\documentclass[article,showpacs]{revtex4-1}
\usepackage{graphics}
\usepackage{dcolumn}
\usepackage{setspace}  
\usepackage{amssymb}
\usepackage{amsmath}
\usepackage{makeidx}
\usepackage{graphicx}
\usepackage{latexsym}
\usepackage{epsfig}
\usepackage{psfrag}
\usepackage{natbib}
\usepackage{rotating}
\usepackage[english]{babel}
\usepackage{tipa}
\usepackage{bm}
\def\TReg {\textsuperscript {\textregistered}}
\usepackage{textcomp}

\begin{document}

\title {EIGENVALUE PROBLEM IN TWO DIMENSIONS FOR AN IRREGULAR BOUNDARY
  II : NEUMANN CONDITION}
\author{S. Panda}
\email{subhasis@cts.iitkgp.ernet.in}
\affiliation{Centre for Theoretical Studies,
Indian Institute of Technology, Kharagpur 721302, India}

\author{S. Chakraborty} 
\email{somdeb.chakraborty@saha.ac.in}
\affiliation {Saha Institute of Nuclear Physics, 1/AF, BidhanNagar,
  Kolkata 700064, India}

\author{S. P. Khastgir}
\email{pratik@phy.iitkgp.ernet.in}
\affiliation{Department of Physics and Meteorology, Indian Institute of Technology, Kharagpur 721302, India}

\begin{abstract}
We formulate a systematic elegant perturbative scheme for
determining the eigenvalues of the Helmholtz equation
$(\nabla^{2}+k^{2})\psi=0$ in two dimensions when the normal
derivative of $\psi$ vanishes on an irregular closed curve. Unique
feature of this method, unlike other perturbation schemes, is that it
does not require a separate formalism to treat degeneracies.
Degenerate states are handled equally elegantly as the non-degenerate
ones. A real parameter, extracted from the parameters defining the
irregular boundary, serves as a perturbation parameter in this scheme
as opposed to earlier schemes where the perturbation parameter is an
artificial one. The efficacy of the proposed scheme is gauged by
calculating the eigenvalues for elliptical and supercircular
boundaries and comparing with the results obtained numerically.  We
also present a simple and interesting semi-empirical formula,
determining the eigenspectrum of the 2D Helmholtz equation with the
Dirichlet or the Neumann condition for a supercircular boundary. A
comparison of the eigenspectrum for several low-lying modes obtained
by employing the formula with the corresponding numerical estimates
shows good agreement for a wide range of the supercircular exponent.
\end{abstract}


\maketitle
\section{INTRODUCTION}
Determination of the eigenspectrum of the Helmholtz equation for
different boundary conditions and geometries has been a long standing
problem both in physics and in engineering. A particular class of the
problem relates to finding out the eigenvalues when the corresponding
eigenfunctions satisfy the Dirichlet or the Neumann condition on the
boundary. In fact, one can also have mixed type of boundary
conditions, although it comes up less frequently in practical
situations than the previous two cases. The problem has been pursued
in different contexts in physics and in engineering. Hence, the
importance of solving the Helmholtz equation for different boundary
conditions can hardly be overemphasized. The standard wave equation
that one encounters once too often in different realms of theoretical
and applied physics reduces to the Helmholtz equation after one
separates out the temporal part. For instance, the canonical example
where one deals with the wave equation is the vibration of membranes
and plates. The problem crops up often even in acoustic and in
mechanical engineering.  Perhaps the best known example of the Neumann
boundary condition is the analysis of the propagation of the TE modes
in a waveguide. Another common example (in 1 dimension) is that of
heat conduction through a rod when energy is supplied at one of its
ends at a constant rate.  Examples of the Dirichlet boundary condition
include the propagation of the TM modes of electromagnetic waves
within a waveguide and the vibration of membranes.  However, in all
these cases simple analytic solutions are available only for a very
restricted class of boundary geometry. In general obtaining a solution
to the problem for an arbitrary boundary can be a formidable
task. However, many physical situations require that we solve the
equation on a general domain.  In particular, the design of waveguides
with a cross-section perturbed from a square may turn out to be useful
in eliminating the losses due to the corners.  The same problem
presents itself, although in a different guise, in quantum mechanics
where one solves the Schr\"{o}dinger equation to find out the energy
states of a particle confined by an infinite potential well in 2
dimensions and enforces the Dirichlet condition on the boundary of the
well. A variant of the same problem is the study of quantum billiards
which, however, involves more subtleties.  Another area which has
witnessed a flurry of activities of late is the study of quantum dots.
The dots are usually taken to have a circular symmetry.  But in
practice that can hardly be guaranteed. There are bound to be small
departures from exact circular symmetry. In such a scenario a very
natural extension is to consider the confining region to be a
supercircle \cite{APL} and investigate the resulting
spectrum. Recently, Bera \textit{et al} \cite{Bera} have proposed a
perturbative approach to the problem where the correction terms were
given by a power series expansion and the method was applied for the
supercircular boundaries. This then calls for a programme to solve the
Helmholtz equation for a general boundary. Chakraborty \textit{et al}
\cite{chak} suggested a general recipe which was in the same spirit of
Bera but furnished the perturbative corrections in a closed form for
the case of the Dirichlet condition.  On a more optimistic note, we
can even attempt to proceed the other way round and expect that a
knowledge of a part of the spectrum may offer an window to probe the
exact shape of the dots. In fact, attempts have already been made
exploring these issues \cite{Lis, Drouvelis, Prb}.  The question
\textit{`Can one hear the shape of a drum?'}  was first posed by Kac
as an inverse eigenvalue problem to the Helmholtz equation \cite{Kac}.
In the context of quantum mechanics this implies whether a knowledge
of the energy states of a particle confined in a 2 dimensional
infinite well is enough to shed light upon the boundary of the well. A
counter-example was constructed by Gordon \textit{et al} \cite{Gordon}
who gave a pair of isospectral but non-congruent domains. However, by
exploiting the \textit{inside-outside} duality of quantum billiard
problems \cite{Doron, Dietz, Berry, Eckmann, Tasaki} it has been shown
recently that the shape of isospectral billiards can indeed be
inferred by studying the exterior Neumann scattering problems
\cite{Revisited}.  Thus, considering its immense applicability, it is
not surprising that the problem of solving the Helmholtz equation
attracts the attention of physicists even today.

However, the problem, both in its classical and quantum incarnations,
is amenable to exact analytic treatment only in some special cases.
Such problems are generally tackled by invoking the method of
separation of variables in a suitably chosen coordinate system. Thus
rectangular and circular boundaries can easily be handled.  The case
of the triangular boundary is more involved \cite{Krishnamurthy}. The
problem of elliptical boundary, though solvable in principle, is quite
non trivial. This is because in elliptical coordinates the separation
of variables leads to the Mathieu equation whose solutions are the
Mathieu functions. Mathieu functions have a 2-parameter dependence
which makes them quite complicated to handle. For any other sort of
boundary we virtually run out of a suitable choice of coordinate
system.  The problem becomes even more untraceable for boundaries
having no simple geometrical shape.  This is a major handicap for an
analytical study of the `irregular boundary' problem.  An alternative
way then is to estimate the eigenvalues by numerical means. In fact,
till now, most of the efforts at finding out the eigenvalues of the
Helmholtz equation for an irregular boundary have been along this
direction \cite{M1, M2, M3, Kuttler, Amore,Amore1, Shaw, Wilson,
  Hettich, Troesch, Kaufman, Vergini, Cohen,Kosztin,
  Robnik,Erwin}. The analytical aspect, however, has not received much
attention except for some occasional attempts.  The problem has been
taken up in different contexts by Rayleigh \cite{Rayleigh}, Fetter and
Walecka \cite{Fetter}, Morse and Feshbach \cite{Morse}, Parker and
Mote \cite{Parker}, Nayfeh \cite{Nayfeh}, Read \cite{Read} and
recently by Wu and Shivakumar \cite{Shiva} and Molinari
\cite{Molinari}. Dubertrand \textit{et al}
\cite{Dub} have employed a similar scheme for studying the propagation
of electromagnetic waves in open dielectric systems.   More recently,
Amore \cite{Amore1} has obtained a systematic approximation to the
ground state energy and the wavefunction on a general domain in a
nonperturbative way. The general spectrum has also been obtained
employing standard perturbative tools.  However, most of these papers
address only the Dirichlet condition.  In this paper we extend the
method of \cite{chak} to encompass the case of the Neumann condition.
This is crucial since it gives us an analytic handle to estimate the
eigenspectrum in an effective and systematic way.  Besides obvious
academic interest this is also necessary if one wishes to consider TE
mode propagation of waveguides through channels of arbitrary
cross-section.  The validity of the method is confirmed by comparing
the analytical results with those obtained numerically for
supercircular and elliptical boundaries. Further, we also propose a
semi-empirical formula, involving only a single parameter, giving the
eigenvalues of the Helmholtz equation for the case of the
supercircular boundary with both type of boundary conditions. To our
knowledge, this is the first instance where a semi-empirical formula
has been put to use to determine the eigenspectrum.  This equips us
with a simple yet effective way to find out the eigenspectrum for the
supercircle.  The validity of the formula is judged by estimating the
energy levels for several low lying modes and comparing them with
numerical calculations.

The paper is organised as follows: In section 2 we set up our general
scheme and in section 3 we apply it to the case of a supercircle and
an ellipse. In section 4 we describe the supercircle and the curious
approximate duality with respect to the supercircular exponent. The
details of the semi-empirical formula for both the Dirichlet and the
Neumann boundary conditions are provided in section 5. A short
conclusion and a discussion of the results are presented in section
6.

\section{PERTURBATION about the equivalent circle}
The homogeneous Helmholtz equation on a 2 dimensional flat simply 
connected surface $\Sigma$ reads,
\begin {equation}
(\nabla^{2}+ k^{2})\psi=0,   \label{he}
\end {equation}
with the Dirichlet condition $\psi=0$ on the boundary $\partial
\Sigma$ or the Neumann condition $\frac{\partial \psi}{\partial n} = 0
$ on $\partial \Sigma$, where $\frac{\partial \psi}{\partial n}$
denotes the derivative along the normal direction to $\partial
\Sigma$. We are interested in finding out $k^{2}(= \omega)$ for the
case of the Neumann condition. Once we are handed any periodic
function $r(\theta)=r(\theta + 2\pi)$ defining the boundary of the
domain $\Sigma$ in flat 2D where we intend to solve the equation we
first construct a circle of radius $R_{0}$ which respects the `equal
area' constraint
\begin{equation}
A_{\Sigma}= \frac{1}{2}\int_{0}^{2\pi}r^{2}(\theta)d\theta=\pi R_{0}^{2} \label{ro}
\end{equation}
where $A_{\Sigma}$ is the area of the domain $\Sigma$.  The next task
then is to Fourier expand $r(\theta)$ about the `equal area' radius
$R_{0}$ at different orders of smallness (denoted by $\lambda$),
\begin{equation}
r(\theta)=R_0\left[1+\sum_{\sigma=1}^{\infty}\lambda^{\sigma}f^{(\sigma)}(\theta)\right], \label{rex}
\end{equation}
where the Fourier series at order $\sigma$ is given as
\begin{equation}
f^{(\sigma)}(\theta)=\sum_{n = 0}^{\infty}\left(C_{n}^{(\sigma)}\cos
n\theta +S_{n}^{(\sigma)} \sin n\theta \right).  \label{gfs}
\end{equation}
This essentially implies that we are seeking a perturbative solution
to the problem in the form of corrections to the solution for a
circular boundary.  While the choice of the smallness parameter
$\lambda$ is certainly dictated to a large extent by the nature of the
boundary we are considering, a certain amount of discretion is also
required on our part to ensure the efficiency of the perturbative
scheme as discussed in detail in the Conclusion of \cite{chak}.  Of
course the Fourier coefficients appearing in ({\bf \ref{gfs}}) are not
all free but constrained by the `equal area' condition,
\begin{align}
\sum_{n=0}^{\infty}\sum_{\nu=1}^{\sigma-1}\left[C_{n}^{(\nu)}C_{n}^{(\sigma-\nu)}+2C_{0}^{(\nu)}C_{0}^{(\sigma-\nu)}+S_{n}^{(\nu)}S_{n}^{(\sigma-\nu)}\right]
= -4C_{0}^{(\sigma)}.
\end{align}
In particular we have,
\begin{align}
C_{0}^{(1)}= &0. \\
4C_{0}^{(2)}= &-\sum_{n=1}^{\infty}\left[C_{n}^{(1)2}+S_{n}^{(1)2}\right]. \label{co2}
\end{align}
Following Rayleigh \cite{Rayleigh} we argue that at the zeroth order
level the eigenvalue $\omega_{0}$ of the system confined by
$r(\theta)$ will be given by that corresponding to the `equal area'
circle of radius $R_{0}$\\
\begin{equation}
\omega_{0}=\frac{\rho^{\prime^{2}}_{l,j}}{R_{0}^{2}}
\ \ \ \ \mathrm{such \ \ that} \ \ \ \ \rho^{\prime}_{l,j} = k^{\prime}_{l,j}R_{0}   \label{gsec}
\end{equation}
with $\rho^{\prime}_{l,j}$ being the $j^{th}$ node of the derivative
of the $l^{th}$ order Bessel function,
i.e. $J_l^{\prime}(\rho^{\prime}_{l,j}) = 0.$ To fine tune the
expression for the eigenvalues the next step will be to incorporate
the correction terms arising due to the deviation of the given
boundary from the equivalent circle.
We do a perturbative expansion of the eigenfunction $\psi$ and the
eigenvalue $\omega$ in terms of the parameter $\lambda$ as
\begin{subequations}
\begin{equation}
\psi=\psi_{0}+\lambda\psi_{1}+\lambda^{2}\psi_{2}+... .\label{ex1}
\end{equation}
\begin{equation}
\omega=\omega_{0}+\lambda \omega_{1}+\lambda^{2}\omega_{2}+...  .\label{ex2}
\end{equation}
\end{subequations}
where $\psi_{0}$ is the unperturbed eigenfunction with corresponding
eigenvalue $\omega_{0}$.  Plugging in these expansions in ({\bf
  \ref{he}}) and collecting separately the coefficients of different
powers of $\lambda$ produces the following set of equations
\begin{subequations}
\begin{align}
\mathcal{O}(0)&:    &\left(\nabla^{2} +\omega_{0}\right)\psi_{0} &= 0
  , \label{0} \\
\mathcal{O}(\lambda)&:   &\left(\nabla^{2}
  +\omega_{0}\right)\psi_{1}&=-\omega_{1}\psi_{0}
  ,\label{1} \\
\mathcal{O}(\lambda^{2})&:   &\left(\nabla^{2} +\omega_{0}\right)\psi_{2}&=-\omega_{1}\psi_{1}-\omega_{2}\psi_{0}.  \label{2}
\end{align}
\end{subequations}
The correction to the eigenfunction at each order will be constrained by
the Neumann boundary condition
\begin{equation}
{\bf \nabla{\psi}}\cdot {\bf n}= 0 \label{nbc}
\end{equation}
where ${\bf n}$ is the normal at the boundary given by $r =
r(\theta).$ Taylor expanding the expression in ({\bf \ref{nbc}})
about $r=R_{0}$ and making use of ({\bf \ref{ex1}}) yields the
constraints at each order,
\begin{subequations} 
\begin{align}
\mathcal{O}(0):\,\,\,\,\,\,\,\, &\psi_{0}^{\prime}(R_{0}) = 0, \label{bc0} \\
\mathcal{O}(\lambda):\,\,\,\,\,\,\,\, &\psi_{1}^{\prime}(R_{0}) + R_{0} f^{(1)}\psi_{0}^{\prime
    \prime}(R_{0})- \frac{1}{R_{0}} \hat{f}^{(1)}
  \hat{\psi_{0}}(R_{0}) = 0,    \label{bc1}\\
\mathcal{O}(\lambda^{2}):\,\,\,\,\,\,\,\, &\psi_{2}^{\prime}(R_{0}) + R_{0}   
f^{(1)} \psi_{1}^{\prime \prime}(R_{0}) + R_{0} f^{(2)}
\psi_{0}^{\prime \prime}(R_{0}) + \frac{1}{2} R_{0}^2 f^{(1)^{2}}
{\psi_{0}}^{\prime \prime \prime}(R_{0}) -
\frac{1}{R_{0}}\hat{f}^{(1)} \hat{\psi_{1}}(R_{0})\nonumber \\ &- \frac{1}{R_{0}}\hat{f}^{(2)} \hat{\psi_{0}}(R_{0}) + \frac{2}{R_{0}}f^{(1)}\hat{f}^{(1)} \hat{\psi_{0}}(R_{0}) = 0 . \label{bc2}
\end{align}
\end{subequations}
In the above expressions prime (${\bf \prime}$) and hat (${\bf
  \hat{\phantom x}}$) denote differentiation with respect to $r$ and
$\theta$ respectively. In the following we discuss separately the
cases $l=0$ and $l \neq 0$.
\subsection{Non-degenerate states ($l=0$)}
For the $l=0$ state we have,
\begin{equation}
\psi_{0}= N J_{0}(\rho) 
\end{equation}
where $J_{0}$ is the $0^{th}$ order Bessel function, and $N$ is a
constant which can be evaluated by normalising the eigenfunction
suitably. $\omega_{0}$ is obtained from ({\bf \ref{gsec}}) with $l=0$,
and an appropriate $j$.\\ To correct the wavefunction upto first order
we need to solve ({\bf \ref{1}}) with the restriction imposed by ({\bf
  \ref{bc1}}).  The most general solution to ({\bf \ref{1}}) is
\begin{equation}
\psi_{1}=\sum_{p=1}^{\infty}(a_{p}\cos p\theta + \bar a_{p}\sin p\theta )J_{p}
+a_{0}J_{0}-\frac{\rho \omega_{1}}{2\omega_{0}}NJ_{1} \label{gfc}
\end{equation}
where the last term is the particular integral.  Constraining this
solution by demanding that it obeys ({\bf \ref{bc1}}) and matching the
coefficients of the sine and the cosine terms separately provides the
coefficients $a_{p}$ and $\bar a_{p}$
\begin{align}
&a_{p} = \rho^{\prime}_{0,j}NC_{p}^{(1)}J_{0}(\rho^{\prime}_{0,j})/J^{\prime}_{p}(\rho^{\prime}_{0,j}),
\\
&\bar a_{p} =
\rho^{\prime}_{0,j}NS_{p}^{(1)}J_{0}(\rho^{\prime}_{0,j})/J^{\prime}_{p}(\rho^{\prime}_{0,j}),
 \ \ \ \ \ \ p\ne 0 
\\ &\omega_{1} = 0 . \label{gse1c}
\end{align}
Simultaneously, we also note that the first correction to the
eigenvalue vanishes for the $l=0$ case.  So any possible correction to
the eigenvalue can only stem from the second or higher order
calculations.  We have one more constant $a_{0}$ to determine which
can be fixed by normalising the corrected eigenfunction over the
region $\Sigma$.  But we will not evaluate it in the present case
since the exact value of $a_{0}$ will not be necessary for the
calculation of the eigenvalues.  \\ The
second order calculation will mimic that of the first order.  Now we
need to solve ({\bf \ref{2}}) with $\omega_{1}$ set equal to 0.  The
most general solution is
\begin{equation}
\psi_{2}=\sum_{p=1}^{\infty}(b_{p}\cos p\theta + \bar b_{p}\sin p\theta )J_{p}
+b_{0}J_{0}-\frac{\rho \omega_{2}}{2\omega_{0}}NJ_{1} \label{gsc}
\end{equation}
where again the last term represents the particular integral.
Imposing the constraint({\bf \ref{bc2}}) on the second order
correction provides expressions for the coefficients $b_{p}$ and $\bar
b_{p}$ together with the second order correction to the eigenvalue,
\begin{subequations} 
\begin{align}
\frac{\omega_{2}}{\omega_{0}} = & -\sum_{n=2}^{\infty}\left[C_{n}^{(1)^{2}}+ S_{n}^{(1)^{2}}\right]\rho^{\prime}_{0,j}
  \frac{J_{n}(\rho^{\prime}_{0,j})}{J_{n}^{\prime}(\rho^{\prime}_{0,j})},  \label{gse2c}
  \\ \nonumber\\
b_{p} = &\frac{\rho^{\prime}_{0,j}J_{0}(\rho^{\prime}_{0,j})}{J^{\prime}_{p}(\rho^{\prime}_{0,j})}[NC^{(2)}_{p}+a_{0}C_{p}^{(1)}]\nonumber \\
&+\frac{N\rho^{\prime}_{0,j}J_{0}(\rho^{\prime}_{0,j})}{4J^{\prime}_{p}(\rho^{\prime}_{0,j})}\sum_{k=1}^{\infty}\left[C_{p+k}^{(1)}C_{k}^{(1)}+S_{p+k}^{(1)}S_{k}^{(1)}+C_{|k-p|}^{(1)}C_{k}^{(1)}+S_{k-p}^{(1)}S_{k}^{(1)}-S_{p-k}^{(1)}
    S_{k}^{(1)}\right]\nonumber
  \\ &+\frac{NJ_{0}(\rho^{\prime}_{0,j})}{2J^{\prime}_{p}(\rho^{\prime}_{0,j})}\sum_{k=1}^{\infty}\left[C_{k}^{(1)}\left\{(\rho^{\prime^{2}}_{0,j}+kp)C_{p+k}^{(1)}+(\rho^{\prime^{2}}_{0,j}-kp)C_{|p-k|}^{(1)}\right\}\right. \nonumber \\
&\left.+S_{k}^{(1)}\left\{(\rho^{\prime^{2}}_{0,j}+kp)S_{p+k}^{(1)}+(\rho^{\prime^{2}}_{0,j}-kp)(S_{k-p}^{(1)}-S_{p-k}^{(1)})\right\}\right]\frac{J_{k}(\rho^{\prime}_{0,j})}{J^{\prime}_{k}(\rho^{\prime}_{0,j})},\\ \nonumber
\end{align}
\begin{align}
\bar{b}_{p} = &\frac{\rho^{\prime}_{0,j}J_{0}(\rho^{\prime}_{0,j})}{J^{\prime}_{p}(\rho^{\prime}_{0,j})}[NS^{(2)}_{p}+a_{0}S_{p}^{(1)}]\nonumber\\
&+\frac{N\rho^{\prime}_{0,j}J_{0}(\rho^{\prime}_{0,j})}{4J^{\prime}_{p}(\rho^{\prime}_{0,j})}\sum_{k=1}^{\infty}\left[S_{p+k}^{(1)}C_{k}^{(1)}-C_{p+k}^{(1)}S_{k}^{(1)}+C_{|k-p|}^{(1)}S_{k}^{(1)}-S_{k-p}^{(1)}C_{k}^{(1)}+S_{p-k}^{(1)}C_{k}^{(1)}\right]\nonumber\\
&-\frac{NJ_{0}(\rho^{\prime}_{0,j})}{2J^{\prime}_{p}(\rho^{\prime}_{0,j})}\sum_{k=1}^{\infty}\left[S_{k}^{(1)}\left\{(\rho^{\prime^{2}}_{0,j}+kp)C_{p+k}^{(1)}-(\rho^{\prime^{2}}_{0,j}-kp)C_{|p-k|}^{(1)}\right\}\right.\nonumber\\
&\left.+C_{k}^{(1)}\left\{(\rho^{\prime^{2}}_{0,j}+kp)S_{p+k}^{(1)}-(\rho^{\prime^{2}}_{0,j}-kp)(S_{k-p}^{(1)}-S_{p-k}^{(1)})\right\}\right]\frac{J_{k}(\rho^{\prime}_{0,j})}{J^{\prime}_{k}(\rho^{\prime}_{0,j})}.
\end{align}
\end{subequations}
So at the second order we expect a nonvanishing correction
to the eigenvalue in general.  The form of $\omega_{2}$ which we have
obtained is completely general and holds for any type of boundary
which departs from a circular one by a small amount.  So for a
particular boundary one only needs to calculate the Fourier
coefficients $C_{n}^{(1)}$ and $S_{n}^{(1)}$ to compute the second order
correction to the eigenvalue.  As before, the remaining constant
$b_{0}$ can be fixed by normalising the wavefunction upto the order of
$\lambda^{2}$.
\subsection{Degenerate states ($l \neq 0$)}
The $l \neq 0$ states come in 2 varieties,
\begin{equation}
\psi_{0}=N_{l}J_{l}(\rho)\left(\begin{array}{c} \cos l\theta \\ \sin l\theta \end{array} \right).
\end{equation}
For these states we shall assume that $S_{n}^{(\sigma)}=0$ for all
$\sigma$.  This is done only to simplify the calculations and will not
mask any crucial physics.  Also for definiteness we prefer to work
with one variety, namely
\begin{equation}
\psi_{0}=N_{l}J_{l}(\rho) \cos l\theta   \ \ \ \ \&
\ \ \ \  \omega_{0}  =\frac{\rho^{\prime^{2}}_{l,j}}{R_{0}^{2}}
\ \ \ \ ;\ \ \ \ J_l^{\prime}(\rho^{\prime}_{l,j}) = 0  
\end{equation}
and just give the result for the other case.  The first correction to
the wavefunction is obtained as a solution to ({\bf \ref{1}}) and
respecting ({\bf \ref{bc1}})
\begin{equation}
\psi_{1}=\sum_{p=0,p\neq l}^{\infty}a_{p}J_{p}\cos p\theta+
\left(a_{l}J_{l}-\frac{\omega_{1}}{\omega_{0}}\frac{\rho}{2}N_{l}J_{l+1}\right
)\cos l\theta.
\end{equation}
As in the case of the $l=0$ case here also the constraint will give
the expressions for $a_{p}$ and $\bar a_{p}$ alongwith the first
correction to the eigenvalue
\begin{subequations} 
\begin{align}
\frac{\omega_{1}}{\omega_{0}}= &
-C_{2l}^{(1)}\left(\frac{\rho^{\prime^{2}}_{l,j}+l^{2}}{\rho^{\prime^{2}}_{l,j}-l^{2}}\right),  \label{ese1cc} \\
a_{0} = & \frac{N_{l} \rho^{\prime}_{l,j}
 C_{l}^{(1)}}{2}\frac{J_{l}(\rho^{\prime}_{l,j})}{J_{0}^{\prime}(\rho^{\prime}_{l,j})}, \\
a_{p} = & \frac{N_{l}J_{l}(\rho^{\prime}_{l,j})}{2
   \rho^{\prime}_{l,j}J_{p}^{\prime}(\rho^{\prime}_{l,j})}
 \left[(\rho^{\prime^{2}}_{l,j}+pl)C_{p+l}^{(1)}
   +(\rho^{\prime^{2}}_{l,j}-pl) C_{|p-l|}^{(1)} \right] ,~~~~~~~~~~~~~~~~~~~~~~\text {for $p\neq 0,l$} 
 \end{align}
\end{subequations}
with $a_{l}$ being obtained from the normalisation condition.
However, we will not evaluate it here.  Unlike in the $l=0$ case now
we expect a non-zero correction to the eigenvalue at the first order
itself.  The second order corrections can be found out in exactly the
same manner, using ({\bf \ref{2}}) \& ({\bf \ref{bc2}}), 
\begin{subequations}
\begin{align}
\psi_{2} = &\sum_{p=0}^{\infty} \left[ b_{p} J_{p} -
  \frac{\omega_{1}}{\omega_{0}} \frac{\rho}{2} a_{p} J_{p+1} \right] \cos p
\theta +\left[\frac{\omega_{1}^2}{\omega_{0}^2} \frac{\rho^2}{4}
  J_{l+2}-\frac{\omega_{2}}{\omega_{0}} J_{l+1}  \right] \frac{\rho N_{l}}{2}\cos
l \theta, \\ 
\frac{\omega_{2}}{\omega_{0}} =  &\frac{1}{2}\frac{\omega_{1}^{2}}{\omega_{0}^{2}}\left(\frac{\rho^{\prime^{2}}_{l,j}-2l^{2}}{\rho^{\prime^{2}}_{l,j}-l^{2}}\right)-\left(\frac{\rho^{\prime^{2}}_{l,j}+l^{2}}{\rho^{\prime^{2}}_{l,j}-l^{2}}\right)C_{2l}^{(2)}-2C_{0}^{(2)}+\left(\frac{\rho^{\prime^{3}}_{l,j}}{\rho^{\prime^{2}}_{l,j}-l^{2}}\right)\frac{J_{0}^{\prime \prime}(\rho^{\prime}_{l,j})}{J_{0}^{\prime}(\rho^{\prime}_{l,j})}C_{l}^{(1)^2}\nonumber
\\&+\frac{1}{4}\left(\frac{\rho^{\prime^{2}}_{l,j}-3l^2}{\rho^{\prime^{2}}_{l,j}-l^2}\right)\sum_{n=1}^{\infty}C_{n}^{(1)}
\left(C_{n+2l}^{(1)} + 2C_{n}^{(1)} + C_{|2l-n|}^{(1)}\right)
\nonumber\\ 
&+\left(\frac{1}{\rho^{\prime^{2}}_{l,j}-l^{2}}\right)\sum_{n=1}^{\infty} C_{n}^{(1)}\left[(nl+2l^{2})C_{n+2l}^{(1)}
  -(nl-2l^{2}) C_{|n-2l|}^{(1)} \right]
\nonumber\\ 
&-\left(\frac{1}{\rho^{\prime^{2}}_{l,j}-l^{2}}\right)\left[\frac{\rho^{\prime^{2}}_{l,j}}{2}\sum_{\substack{n=1\\ n\neq
    l}}^{\infty}\left(C_{n+l}^{(1)}+ C_{|n-l|}^{(1)}\right)^{2}+\frac{1}{2}\sum_{\substack{n=1\\ n\neq
    l}}^{\infty} nl\left(C_{n+l}^{(1)^{2}}-C_{|n-l|}^{(1)^{2}}\right)\right. \nonumber
  \\ 
&\left.+\frac{1}{2\rho^{\prime}_{l,j}} \sum_{\substack{n=1\\ n\neq
    l}}^{\infty}\left\{\left[(\rho^{\prime^{2}}_{l,j}+nl)C_{n+l}^{(1)}+(\rho^{\prime^{2}}_{l,j}-nl)C_{|n-l|}^{(1)}  \right]^{2} \frac{J_{n}(\rho^{\prime}_{l,j})}{J_{n}^{\prime}(\rho^{\prime}_{l,j})}\right\}\right]. \label{ese2cc} 
\end{align}
\end{subequations}
The constants $b_{m}$ can also be determined as mentioned earlier. The
equation for $\psi_{2}$ ({\bf \ref{2}}) also involves $\psi_{1}$ in
one of the inhomogeneous terms. So it may seem that we need to know
the coefficient $a_{l}$ (which we have not found out here) to
determine the form of $\psi_{2}$. However, it turns out that the terms
involving $a_{l}$ cancel out amongst themselves so that we really do
not need to explicitly evaluate it.  For the other variety,
\begin{equation}
\psi_{0}=N_{l}J_{l}(\rho)\sin l\theta
\end{equation}
we can proceed in an analogous manner. However, for the sake of
completeness we give only the final results here,
\begin{subequations} 
\begin{align}
&\psi_{1}=\sum_{p=1, p\neq l}^{\infty}\bar a_{p}J_{p}\sin p\theta+
\left(\bar
a_{l}J_{l}-\frac{\omega_{1}}{\omega_{0}}\frac{\rho}{2}N_{l}J_{l+1}\right
)\sin l\theta, \\
&\frac{\omega_{1}}{\omega_{0}}=C_{2l}^{(1)}\left(\frac{\rho^{\prime^{2}}_{l,j}+l^{2}}{\rho^{\prime^{2}}_{l,j}-l^{2}}\right), \label{ese1cs} \\
&\bar{a}_{p}=\frac{N_{l}J_{l}(\rho^{\prime}_{l,j})}{2
   \rho^{\prime}_{l,j}J_{p}^{\prime}(\rho^{\prime}_{l,j})}
 \left[(\rho^{\prime^{2}}_{l,j}-pl) C_{|p-l|}^{(1)}-
   (\rho^{\prime^{2}}_{l,j}+pl)C_{p+l}^{(1)}\right],~~~~~~~~~~~~~~~~~~~~~~~~~~~~~~~~\text {for $p\neq l$}
\end{align}
and
\begin{align}
\psi_{2} = &\sum_{p=1}^{\infty} \left[ \bar{b_{p}} J_{p} -
    \frac{\omega_{1}}{\omega_{0}} \frac{\rho}{2} \bar{a_{p}} J_{p+1}
    \right] \sin p \theta
  +\left[\frac{\omega_{1}^2}{\omega_{0}^2}
    \frac{\rho^2}{4} J_{l+2}-\frac{\omega_{2}}{\omega_{0}} J_{l+1}
    \right]\frac{\rho N_{l}}{2} \sin l
  \theta,\\ \nonumber\\
\frac{\omega_{2}}{\omega_{0}} = & \frac{1}{2}\frac{\omega_{1}^{2}}{\omega_{0}^{2}}\left(\frac{\rho^{\prime^{2}}_{l,j}-2l^{2}}{\rho^{\prime^{2}}_{l,j}-l^{2}}\right)+\left(\frac{\rho^{\prime^{2}}_{l,j}+l^{2}}{\rho^{\prime^{2}}_{l,j}-l^{2}}\right)C_{2l}^{(2)}-2C_{0}^{(2)}\nonumber\\ 
&+\frac{1}{4}\left(\frac{\rho^{\prime^{2}}_{l,j}-3l^2}{\rho^{\prime^{2}}_{l,j}-l^2}\right)\sum_{n=1}^{\infty}C_{n}^{(1)}\left(2C_{n}^{(1)}-C_{n+2l}^{(1)}-C_{|2l-n|}^{(1)}\right)\nonumber\\ 
&-\left(\frac{1}{\rho^{\prime^{2}}_{l,j}-l^{2}}\right)\sum_{n=1}^{\infty}C_{n}^{(1)}\left[(nl+2l^{2})C_{n+2l}^{(1)}-(nl-2l^{2})C_{|n-2l|}^{(1)}\right]\nonumber\\
&-\left(\frac{1}{\rho^{\prime^{2}}_{l,j}-l^{2}}\right)
\left[\frac{\rho^{\prime^{2}}_{l,j}}{2}\sum_{\substack{n=1\\ n\neq
        l}}^{\infty}\left(C_{n+l}^{(1)}-
    C_{|n-l|}^{(1)}\right)^{2}+\frac{1}{2}\sum_{\substack{n=1\\ n\neq
        l}}^{\infty} nl\left(C_{n+l}^{(1)^{2}}-
    C_{|n-l|}^{(1)^{2}}\right)\right. \nonumber  \\ 
&\left.+\frac{1}{2\rho^{\prime}_{l,j}} \sum_{\substack{n=1\\ n\neq l}}^{\infty}\left[(\rho^{\prime^{2}}_{l,j}+nl)C_{n+l}^{(1)}-(\rho^{\prime^{2}}_{l,j}-nl)C_{|n-l|}^{(1)}\right]^{2}\frac{J_{n}(\rho^{\prime}_{l,j})}{J_{n}^{\prime}(\rho^{\prime}_{l,j})}\right] \label{ese2cs}
\end{align}
\end{subequations}
\section {APPLICATION to simple cases}
\subsection{Introduction}
In the preceding section we have formulated the general formalism in
detail.  In this section we put this formalism to test by determining
the spectrum when the boundary is supercircular or elliptical. Direct
comparison with numerical results are also made.  We have solved the
equation numerically, using the Partial Differential Equation
Toolbox${\rm ^{TM}}$ of MATLAB$\TReg $ by defining our supercircular
or elliptical boundary into it.
\subsection{Supercircular boundary}
\begin{figure}
\centering 
\rotatebox{0}{\scalebox{1}{\includegraphics{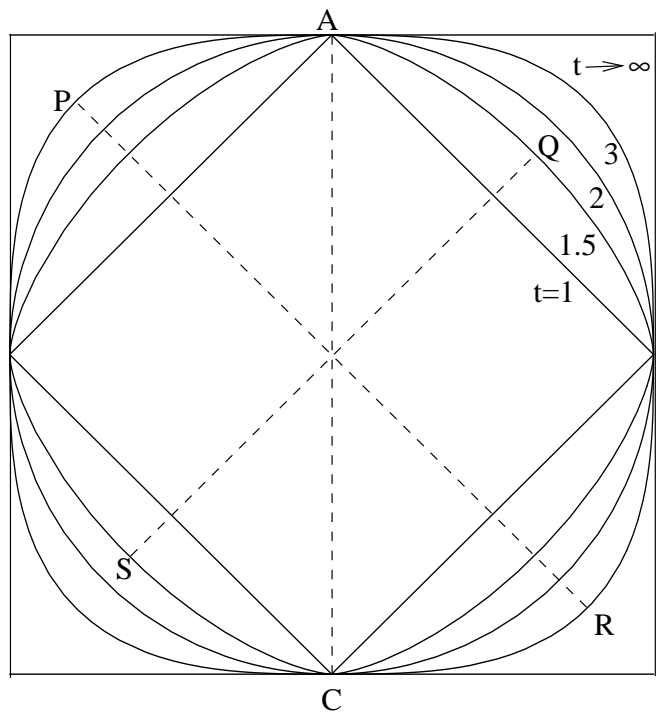}}}
\caption{Shape of the supercircle for different values of
  $t$.}
\label{fig:I}
\end{figure}

\begin{figure}
\centering
\psfrag{Energy (E)}[c][c][1][0]{{\bf {\LARGE Energy ($\omega$)}}}
\rotatebox{0}{\scalebox{0.6}{\includegraphics{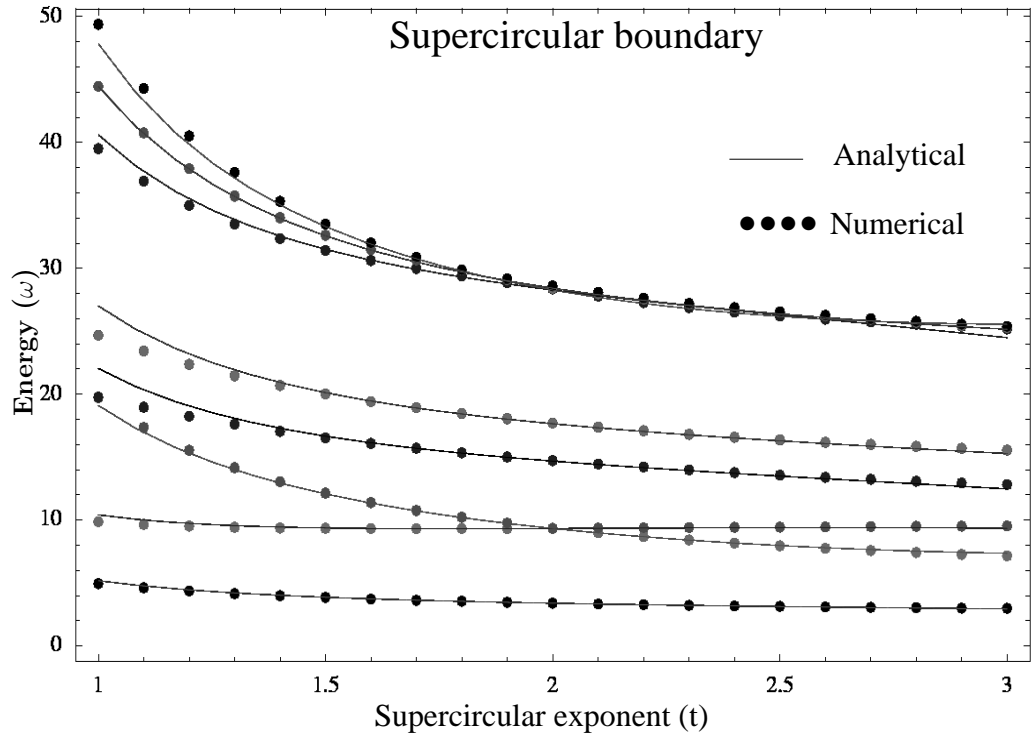}}}
\caption{Comparison of the eigenvalues obtained numerically and
  analytically for a supercircular boundary (in units of
  $\frac{1}{R_{o}^{2}}$) with Neumann condition for the first 11
  states.}
\label{fig:II}
\end{figure}

\begin{figure}
\centering
\psfrag{Energy (E)}[c][c][1][0]{{\bf {\LARGE Energy ($\omega$)}}}
\rotatebox{0}{\scalebox{0.6}{\includegraphics{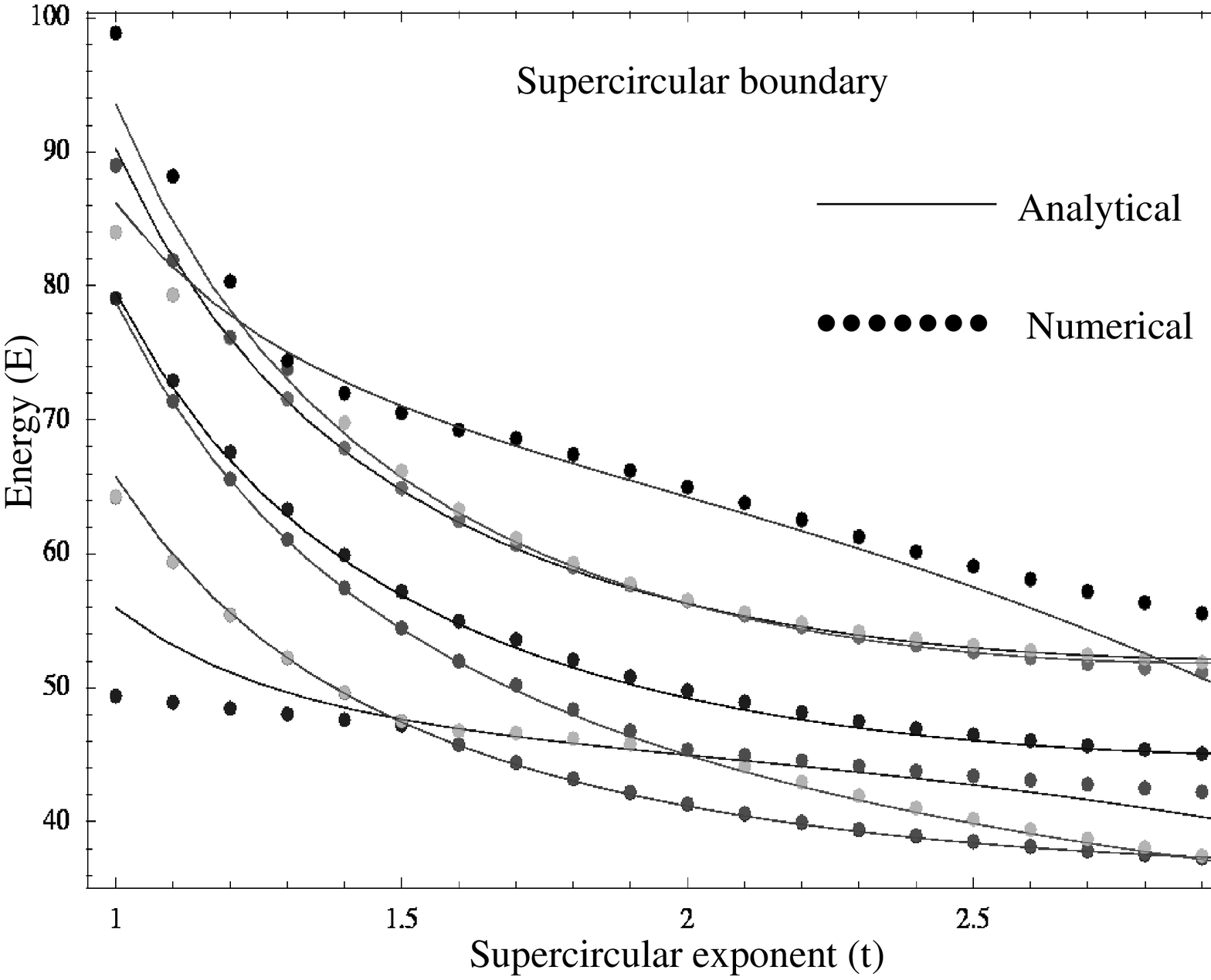}}}
\caption{Comparison of the eigenvalues obtained numerically and
  analytically for a supercircular boundary (in units of
  $\frac{1}{R_{o}^{2}}$) with Neumann condition for the states 12 to
  20.}
\label{fig:III}
\end{figure}

\begin{figure}
\centering
\psfrag{Energy (E)}[c][c][1][0]{{\bf {\LARGE Energy ($\omega$)}}}
\rotatebox{0}{\scalebox{0.6}{\includegraphics{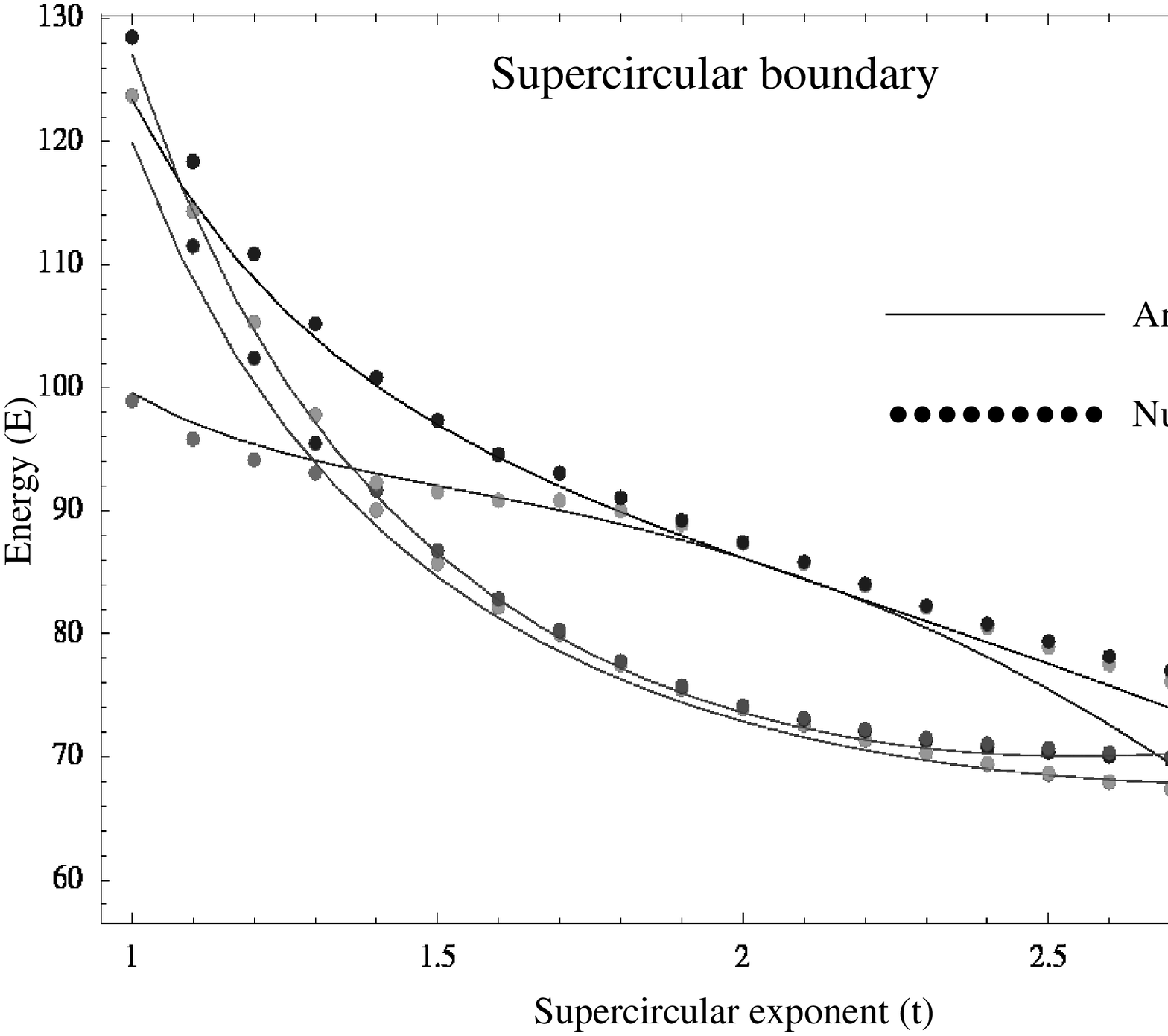}}}
\caption{Comparison of the eigenvalues obtained numerically and
  analytically for a supercircular boundary (in units of
  $\frac{1}{R_{o}^{2}}$) with Neumann condition for the states 21 to
  26.}
\label{fig:IV}
\end{figure}

Supercircle is a special case of a superellipse \cite{Gardner} whose
equation is given by the Lam\'{e} equation,
\begin{equation}
\frac{|x|^{t}}{a^{t}}+ \frac{|y|^{t}}{b^{t}}=1.  \label{sc}
\end{equation}
with $t>0$ and rational, $a$ and $b$ being positive real numbers. In
the literature they are also known as Lam\'{e} curves or Lam\'{e} ovals
\cite{Gridgeman}. Superellipses can be parametrically described as,
\begin{align}
x =&a\cos^{2/t}(u),\\
y =&b\sin^{2/t}(u).
\end{align}
A supercircle will obviously correspond to setting $a=b$.  For each
value of $`t$' - the supercircular exponent we obtain a different curve.
The shapes of the supercircles for different values of $t$ are shown
in FIG.\ref{fig:I}.  Evidently, $t=2$ corresponds to a circle whereas
$t=1$ corresponds to a square with its sides rotated by an angle of
$\frac{\pi}{4}$. The case $t \rightarrow \infty$ also corresponds to a
square with its sides parallel to the axes.  In these limiting cases
the Helmholtz equation can be trivially solved for both Dirichlet and
Neumann type boundary conditions without invoking any special
techniques.  We will, however, be interested in a general value of
$t$.  More specifically, in the spirit of the perturbative formalism
we have developed we would like to find the spectrum when $t$ slightly
deviates from 2. The problem was addressed in \cite{chak} for the
Dirichlet condition. In this paper our aim will be to solve the same
problem for the Neumann type boundary condition.  When $t>2$ we shall
take only the real positive values of $ \cos^{\frac{2}{t}}(u)$ and
$\sin^{\frac{2}{t}}(u) $. Moreover, while calculating the
eigenfunction we will confine our attention to the region $0 \leqslant
u \leqslant \frac{\pi}{2} $ and exploit the symmetry of the region to
continue to the other quadrants. In polar coordinates the equation for
the supercircle is,
\begin{equation}
r=\frac{a}{(\cos^{t}\theta+\sin^{t}\theta)^{1/t}}
\end{equation}
and the radius of the equal area circle is,
\begin{equation}
R_{0}=a\sqrt{\frac{2}{t\pi}}\frac{[\Gamma(\frac{1}{t})]}{\sqrt{[\Gamma(\frac{2}{t})]}}.
\end{equation}
We define the deformation parameter to be $\lambda = t-2$.  The
Fourier expansion of $r(\theta)$ is given as
\begin{equation}
r=R_{0}\left[ 1+ \lambda\sum_{n=1}^{\infty}C_{4n}^{(1)}\cos 4n\theta + \lambda^{2}\sum_{n=0}^{\infty} C_{4n}^{(2)}\cos 4n\theta \right]
\end{equation}
with
$$C_{4n}^{(1)}=-\frac{1}{4n(4n^2-1)}.$$ $C_{0}^{(2)}$ is given by
equation ({\bf \ref{co2}}) to be -0.0017552 and
$C_{4}^{(2)}=\frac{1}{32}\left(\frac{3\pi^{2}}{8}-\frac{23}{9}\right)$=
0.0357983 \cite{chak}.
\\ A knowledge of these few coefficients is sufficient to determine the
eigenvalue corrections to second order.  In FIG.\ref{fig:II},
FIG.\ref{fig:III} and FIG.\ref{fig:IV} we have plotted the first few
levels (continuous line) using our scheme and compared with the
numerical results (discrete points).  We have varied the deformation
parameter in the range $-1 \leqslant \lambda \leqslant 1$.  We find
that the analytical results tally with the numerical ones for small
perturbations around the circle, i.e. for small value of
$|\lambda|$. Within the range $1.5 \leq t \leq 2.5$ the matching is
quite impressive. As we increase the deviation the agreement becomes
poor which is not unexpected. 
Even then for a distortion as large as $t=3$ the agreement with
numerical results for many of the low lying modes 
is quite encouraging indeed.

\subsection {Elliptical boundary} 
\begin{figure}
\centering
\psfrag{Energy (E)}[c][c][1][0]{{\bf {\LARGE Energy ($\omega$)}}}
\rotatebox{0}{\scalebox{0.6}{\includegraphics{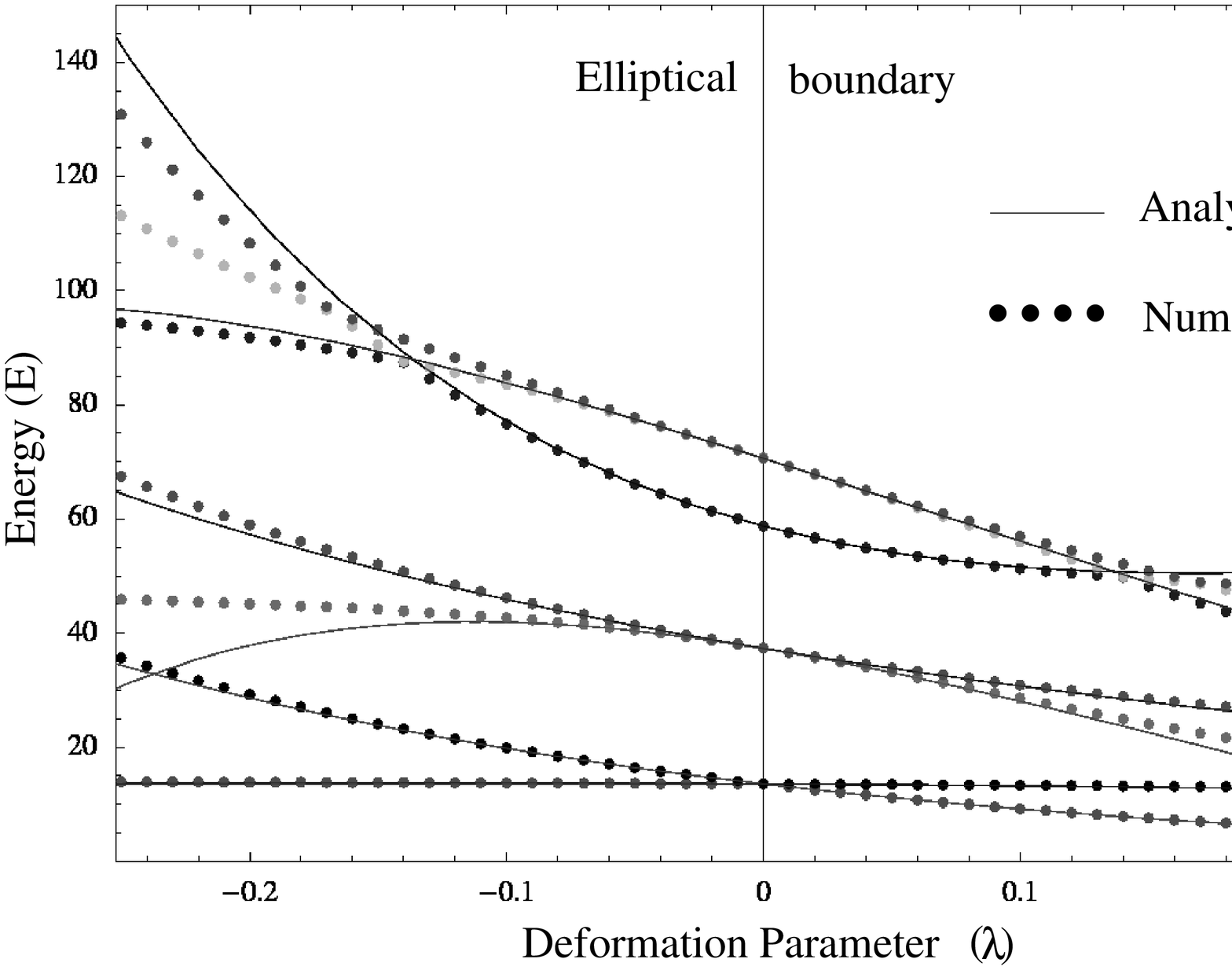}}}
\caption{Comparison of the eigenvalues obtained numerically and
  analytically for an elliptical boundary (in units of
  $\frac{1}{R_{o}^{2}}$) with Neumann condition for the first 7 states.}
\label{fig:V}
\end{figure}

\begin{figure}
\centering
\psfrag{Energy(E)}[c][c][1][0]{{\bf {\LARGE Energy ($\omega$)}}}
\rotatebox{0}{\scalebox{0.6}{\includegraphics{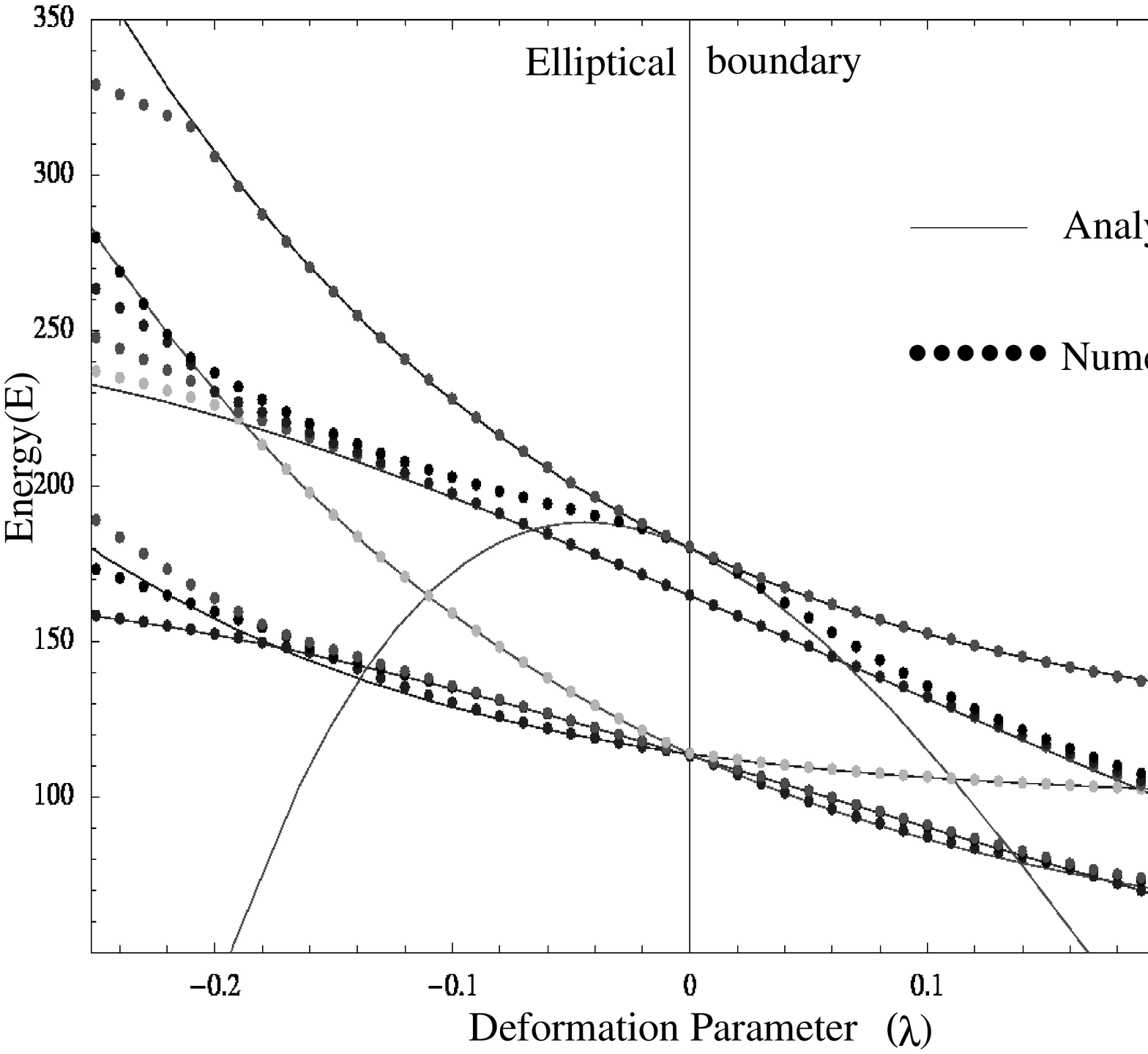}}}
\caption{Comparison of the eigenvalues obtained numerically and
  analytically for an elliptical boundary (in units of
  $\frac{1}{R_{o}^{2}}$) with Neumann condition for the states 8 to
  15.}
\label{fig:VI}
\end{figure}

As a second example we determine the eigenvalues when the boundary is
an ellipse. The case of elliptical boundary has been studied
extensively in the literature because it is one of the few cases where
the Helmholtz equation can be solved exactly.  Even then it is an
arduous task since it involves the Mathieu functions and very often
one has to depend upon numerical methods \cite{Wilson, Hettich,
  Troesch}.  Recently, Wu and Shivakumar \cite{Shiva} attempted an
analytical solution to the problem. Here we will use our perturbative
algorithm to solve the problem for the Neumann type boundary
condition. The polar equation for an ellipse with semiaxes $a$ and $b$
is
\begin{equation}
r=\frac{b}{\sqrt{1-(1-\frac{b^2}{a^2})\cos^2\theta}}.  \label{eqe1}
\end{equation}
We will choose,
\begin{equation}
\lambda=\frac{a-b}{a+b}
\end{equation}
as our deformation parameter although the eccentricity might seem to
be a natural choice.  This is one instance where one needs to exercise
some discretion to ensure the effeciency of the perturbation theory
and enhance the precision of the analytical predictions (see the
Conclusion in \cite{chak}).  ({\bf \ref{eqe1}}) can be recast as,
\begin{equation}
r=R_0[1+\lambda \cos 2\theta -\frac{1}{4}\lambda^2 +
\frac{3}{4}\lambda^2 \cos 4\theta + {\rm O}(\lambda^{3})+...]   .     \label{eqe2}
\end{equation}
A comparison with our general Fourier series yields the Fourier
coefficients relevant to our purpose, $C_2^{(1)}=1$,
$C_{0}^{(2)}=-\frac{1}{4}$ and $C_{4}^{(2)}=\frac{3}{4}$.\\
The results for the elliptical boundary are shown
in FIG.\ref{fig:V} and FIG.\ref{fig:VI}. The deformation parameter
varies from -0.25 to +0.25.  Here also the analytical predictions
tally with the numerical calculations for a fairly wide range of
$\lambda$. The only notable exceptions are the cases where the levels
repel each other. For those cases also the formalism works well for
one of the levels ($J_{l}\sin l\theta$) involved. However for the
other level ($J_{l}\cos l\theta$) it falters.
For the other cases, the matching between analytical and numerical
results is quite pronounced between $-0.15 \leq \lambda \leq 0.15$.  For
a larger deformation the agreement becomes less satisfactory.  It
should also be mentioned that in some of the higher states our
numerical estimates may not be very accurate. For example it is seen
that while the analytical results forecast exact degeneracy of some of
the higher states throughout the entire range of $\lambda$, the
numerical technique that we have used reflects that degeneracy only in
a certain range of $\lambda$.  Outside that range the degeneracy is
lifted. So the apparent disparity between the analytical and the
numerical estimates has got to do with the limitation of the numerical
scheme we have employed rather than the failure of the perturbative
formulation for the higher modes.
 Perhaps one can employ some other numerical techniques and
see whether the agreement improves for the higher states and  whether 
the degeneracy is preserved.
\section{ THE supercircular duality}
\begin{figure}
\centering
\rotatebox{0}{\scalebox{0.6}{\includegraphics{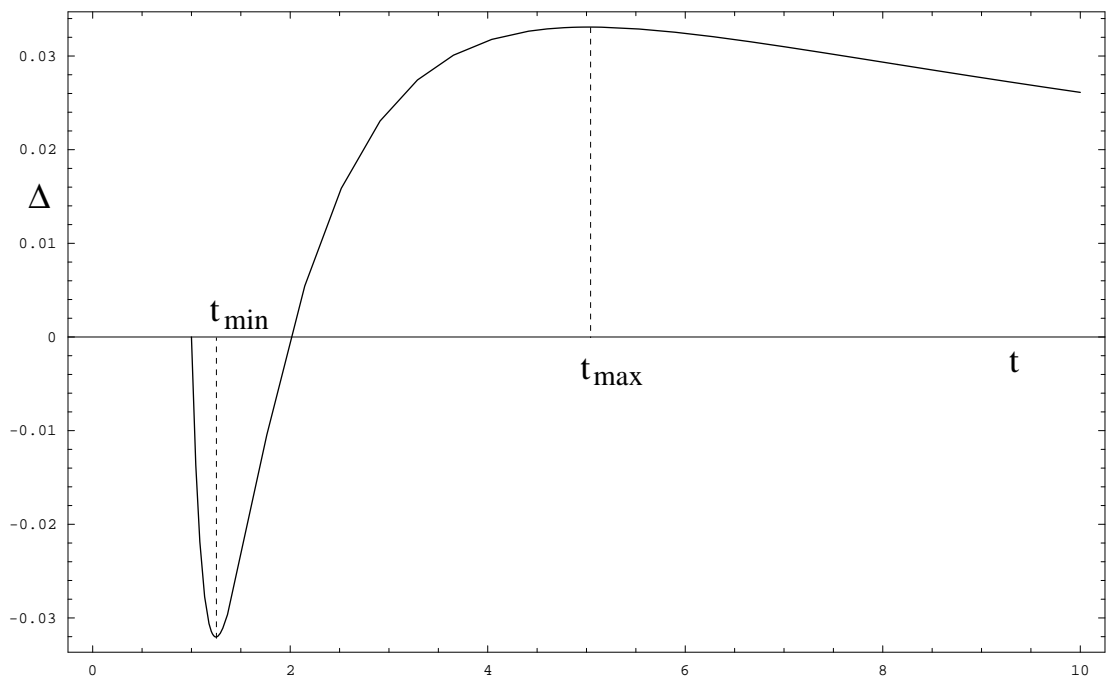}}}
\caption{Justification of the duality relation.}
\label{fig:VII}
\end{figure}

In this section we discuss a very interesting duality for the case of
a supercircle which will prove to be immensely helpful when we
construct the semi-empirical formula in the next section.  Without
loss of generality henceforth we choose $a=1$ in the equation of a
supercircle.  We also restrict $t \ge 1$, as before, so that we always
have a convex curve. The area of the supercircle is given by,
\begin{equation}
A=\frac{2}{t}\frac{[\Gamma(\frac{1}{t})]^{2}}
{[\Gamma(\frac{2}{t})]}. \label{area}
\end{equation}
We show that for indices $t$ and
$t'$ related by the following \cite{Bera},
\begin{equation}
 \frac{1}{t}+ \frac{1}{t'}=1 ; \  \ 1 \le t \le 2  \ \ \mathrm{and} 
\ \ \infty \ge t' \ge 2, \label{dual} 
 \end{equation}
the shapes of the supercircles are almost identical. Clearly, a circle
is self dual since for a circle $t=t'=2$. On the other hand, the 2
squares $-$ one corresponding to $t=1$ and the other corresponding to
$t' = \infty$ $-$ are also dual to each other. But the areas of the 2
squares are different and differ just by a scale factor. However, the
duality we define in this way is not exact. In fact, the two aforesaid
cases are the only ones where we have an exact duality and our naive
argument of scaling the area (the scale factor being unity for the
self-dual case) holds true.  Nevertheless, it is an enticing prospect
to explore how far this simple scaling argument can give us an
estimate in the other cases. For our purpose it is convenient to
recast (using the properties of the Gamma function, $\Gamma(2z) =
\frac{2^{2z-1}}{\sqrt \pi} \Gamma(z) \Gamma(z+1/2)$) the area $A$ of
the supercircle (with $1 \le t \le 2 $) as
\begin{equation}
A(t) = \frac{4\sqrt\pi\Gamma(1+\frac{1}{t})}{2^{2/t}
  \Gamma(\frac{1}{2}
+\frac{1}{t})}.
\end{equation}
For $ 2 \le t' \le \infty $ the area when expressed in terms of
its dual exponent $t$ defined by the relation ({\bf \ref{dual}}) is,
\begin{equation}
A(t')
=A\left(\frac{t}{t-1}\right)= \frac{4 \sqrt\pi \Gamma(2-\frac{1}{t})}
{2^{(2-\frac{2}{t})}
  \Gamma(\frac{3}{2}-\frac{1}{t})}.
\end{equation}
In FIG.\ref{fig:I} we have shown the major axes, AC and PR, for
$t=1.5$ and its dual parameter $t'=3$. The corresponding minor axes
are QS and AC respectively. In general, the ratio, $\alpha$, of the
major axis to the minor axis (for example AC/QS) for $1<t<2$ is
$2^{(\frac{1}{t} - \frac{1}{2})}$. The ratio, $\alpha'$, of the major
axis to the minor axis for $2<t'< \infty$ (for example PR/AC) is
$2^{(\frac{1}{2} - \frac{1}{t'})}$. Now it is easy to see that the
above two ratios are same, i.e. $\alpha=\alpha'$, if $t$ and $t'$
satisfy the duality relation ({\bf \ref{dual}}). $\alpha$ also measures
the scale (the ratio of two major (or minor) axes, i.e. PR/AC (=
AC/QS)) through which the supercircle is inflated under the duality
transformation ({\bf \ref{dual}}). In case of exact duality the area
should scale by $\alpha^{2}$, i.e. $A(t')= \alpha^{2} A(t)$. In
FIG.\ref{fig:VII} we have plotted $\Delta = \frac{\alpha^2
  A(t)-A(t')}{A(t')}$ against $t$ to see 
whether such a scaling law gives a good estimate of the area
for the dual figure. The maximum aberration from 
the scaling law is seen to  take place
at two points viz. $t_{min} \approx 1.2485$ and $t_{max} \approx
5.0242$ and it is obvious that $t_{min}$ and $t_{max}$ satisfy the
duality relation ({\bf \ref{dual}}). The deviation is zero at $t=1$ and
$t=2$ thereby vindicating our previous assertion that these 2 cases
exhibit exact duality. For the other cases the deviation is very small
(within $3.3\%$) and by virtue of that we can safely exploit the duality relation
for extracting the eigenvalues in the domain $t\geq 2$ once the results 
are known for $1 \leq t\leq 2$.
\section{THE semi-empirical formula for supercircular boundary}
In this section we propose a semi-empirical formula for determining
the eigenspectrum for the case of the supercircular boundary for both
type of boundary conditions.  In the following we shall consider the
Dirichlet and the Neumann boundary conditions separately. The case of
the Dirichlet condition is discussed in detail. The Neumann condition
can be similarly treated and hence only briefly dwelt upon.
\subsection{Dirichlet boundary condition}
\begin{figure}
\centering
\psfrag{Energy (E)}[c][c][1][0]{{\bf {\LARGE Energy ($\omega$)}}}
\rotatebox{0}{\scalebox{0.7}{\includegraphics{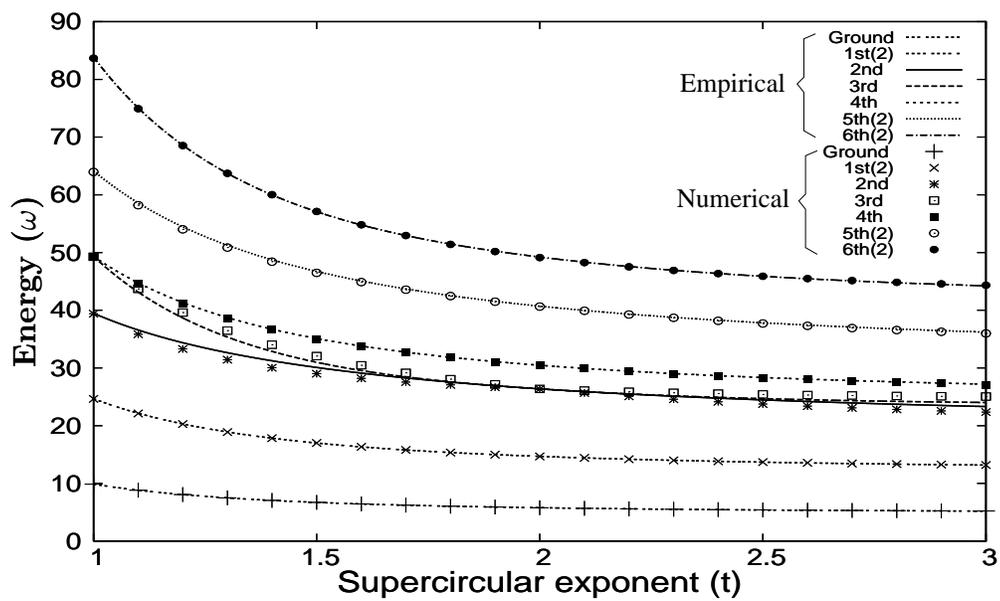}}}
\caption{Comparison of the eigenvalues obtained numerically and
  empirically for a supercircular boundary with Dirichlet condition
  for the first 10 states.}
\label{fig:VIII}
\end{figure}

\begin{figure}
\centering
\psfrag{Energy(E)}[c][c][1][0]{{\bf {\LARGE Energy ($\omega$)}}}
\rotatebox{0}{\scalebox{0.6}{\includegraphics{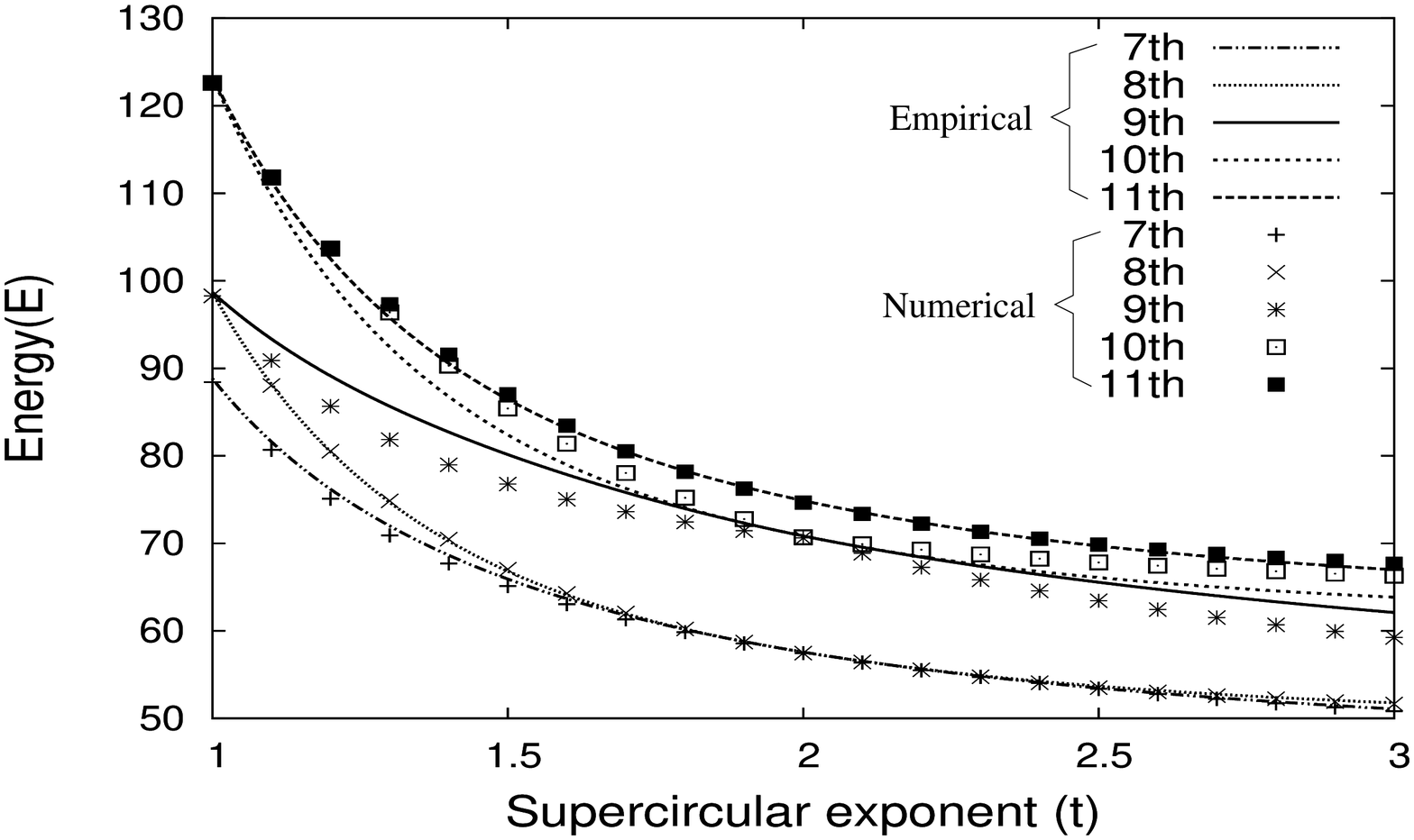}}}
\caption{Comparison of the eigenvalues obtained numerically and
  empirically for a supercircular boundary with Dirichlet condition
  for the states 11 to 15.}
\label{fig:IX}
\end{figure}

\begin{figure}
\centering
\psfrag{Energy(E)}[c][c][1][0]{{\bf {\LARGE Energy ($\omega$)}}}
\rotatebox{0}{\scalebox{0.6}{\includegraphics{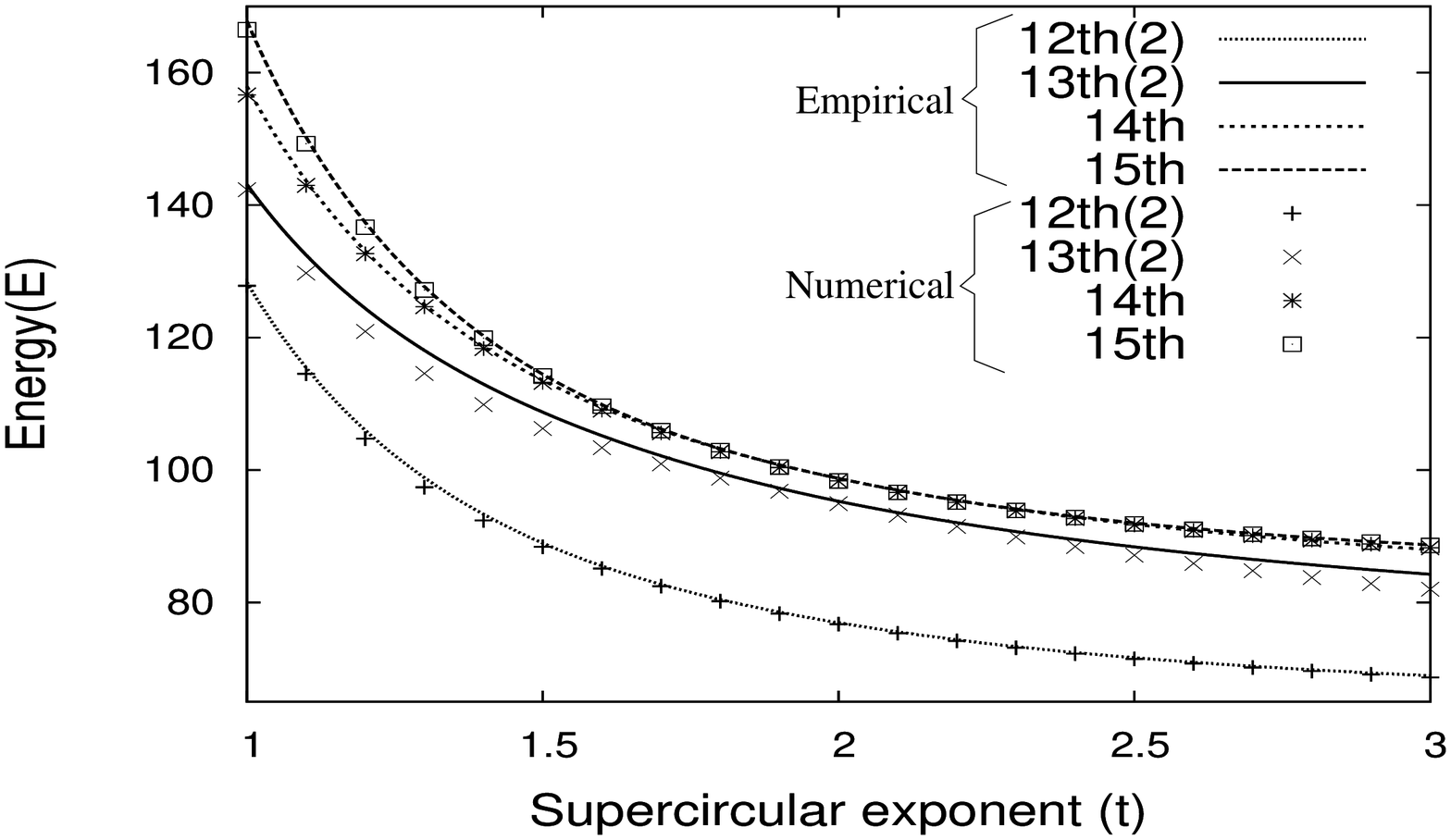}}}
\caption{Comparison of the eigenvalues obtained numerically
  and empirically for a supercircular boundary with Dirichlet
  condition for the states 16 to 21.}
\label{fig:X}
\end{figure}

Given a supercircle of area $A$ we consider a square of length $L$ and
a circle of radius $R$ and each having an area $A$. The length of each
side of the `equal area' square is,
\begin{equation}
L=\sqrt{\frac{2}{t}}\frac{[\Gamma(\frac{1}{t})]}
{\sqrt{[\Gamma(\frac{2}{t})]}}, \label{sqr1}
\end{equation}
and the radius of the `equal area' circle is,
\begin{equation}
R=\sqrt{\frac{2}{t\pi}}\frac{[\Gamma(\frac{1}{t})]}
{\sqrt{[\Gamma(\frac{2}{t})]}}. \label{cir1}
\end{equation}
The solution to ({\bf \ref{he}}), when the boundary is a
square ($S$) of side $L$ is,
\begin {subequations}
 \begin{equation}
  \psi=\frac{2}{L} \sin\left(\frac{n_{x}\pi x}{L}\right)\sin
\left(\frac{n_{y}\pi y}{L}\right),
 \end{equation}
and hence the eigenvalue,
\begin{equation}
 \omega_{s}= k_{s}^{2}= (n_{x}^{2} + n_{y}^{2})\frac{\pi^{2}}{L^{2}}
\end{equation} \label{sqr}
\end {subequations}
where $n_{x}$ and $n_{y}$ are positive integers. It is then easy to
see that $\psi=0$ on $\partial S$ (i.e. on $x = 0 $, $x = L$, $y = 0$ and
$y = L$).
When the boundary is a circle ($C$) of radius $R$, the solution is,
\begin{subequations}
\begin{equation}
 \psi=\left(\begin{array}{c} \frac{1}{\sqrt{\pi} R J_{1}(\rho_{0,j})} J_{0}(\rho) \\ \frac{\sqrt{2}}{\sqrt{\pi} R J'_l(\rho_{l,j})} J_{l}(\rho)
\left(\begin{array}{c} \cos l\theta \\ \sin l\theta \end{array}
\right)   \end{array} \right),
\end{equation}
with $\rho_{l,j}= k_{c}R$ and
\begin{equation}
 \omega_{c} = k_{c}^{2}=\frac{\rho_{l,j}^{2}}{R^{2}}
\end{equation} \label{cir}
\end{subequations}
where $\rho_{l,j}$ is the $j^{th}$ zero of the $l^{th}$ order Bessel
function, $J_{l}(\rho)$. Now to build up a semi-empirical formula for
the case of a supercircle we will attempt to find an expression for
the eigenvalue $\omega(t)$ in the range $1 \le t \le 2$. Further, the
duality discussed in the preceding section will allow us to extend our
results to $2 \le t \le \infty$ making the whole domain as $1 \le t
\le \infty$. It is easy to figure out from FIG.\ref{fig:I} that in the
range $1< t < 2$ the boundary has a shape which is an intermediate
between a circle and a square. So it makes sense to try an expression
of the form,
\begin{equation}
\omega(t)= f(t)\omega_{s}+(1-f(t))\omega_{c},
\end{equation}
i.e. we take an average of $\omega_{s}$ and $\omega_{c}$ with suitable
weight factors. We note that all the eigenvalues are calculated for
boundaries enclosing an equal area given by ({\bf \ref{area}}). Hence
the $L$ and the $R$ in the equation ({\bf \ref{sqr}}) and ({\bf
  \ref{cir}}) are identified with the ones appearing in the equation
({\bf \ref{sqr1}}) and ({\bf \ref{cir1}}) respectively. To find an
explicit form of $f(t)$ we note that in the limiting cases $t=1$ and
$t=2$ the semi-empirical formula must reduce to,
\begin{equation}
\omega(1)= \omega_{s} \ \ \mathrm{and} \ \ \omega(2)= \omega_{c}.
\end{equation}
In other words, in such limits, we must have,
\begin{equation} 
f(1)=1\ \  \mathrm{and} \ \ f(2)=0. \label{lower1}
\end{equation} 
Clearly, there can be infinitely many choices of $f(t)$ fulfilling the
above criteria ({\bf \ref{lower1}}). The simplest choice is $f(t)=
2-t$. Unfortunately the eigenvalues calculated with this choice does
not match very well with the corresponding numerical results. So an
exact form of these weight factors cannot be found out from the above
arguments only and is best obtained through a `trial and error'
method. We have explored different choices of $f(t)$, calculating the
eigenvalues and comparing them with the numerical results. As the next
plausible candidate we take $f(t) = (2-t)^{n}$ and try to estimate the
best $n$ using numerical results. Finally we have settled for the
choice, $n=3$, so that
\begin{equation}
f(t)=(2-t)^{3}.  \label{f}
\end{equation}
So the final expression for the eigenvalue for a supercircle with an
exponent $t$ $(1 \le t \le 2)$ is furnished through the following
expression:
\begin{equation}
\omega(t)=\left[(2-t)^{3}{\cal{E}}_{s}(n_x,n_y)+[1-(2-t)^{3}]{\cal{E}}_{c}
(\rho_{l,j})\right]\frac{
  t \Gamma(\frac{2}{t})}{2[\Gamma(\frac{1}{t})]^{2}}, \label{first}
\end{equation}
where ${\cal E}_s (=L^2\omega_s)$ and ${\cal E}_c (=\pi R^2\omega_c)$
are the eigenvalues for a unit area square and a unit area circle
respectively. For a particular excited mode $n_x, n_y$ and
$\rho_{l,j}$ have to be chosen properly. In Table \ref{table:1} we
have shown the exact mapping between the parameters $n_x$ and $n_y$ of
${\cal E}_s$ and $\rho_{l,j}$ of ${\cal E}_c$ for the first 10
states. The eigenvalues (including the degeneracies) ${\cal E}_s$ and
${\cal E}_c$ are arranged in increasing order of magnitude and are
then matched one to one. Note that all the states with $l \neq 0$ for
the circle are doubly degenerate and hence are written twice. Each
pair of $n_{x}$ and $ n_{y}$ corresponds to one $\rho_{l,j}$. Here we
have excluded the trivial case $n_{x} = n_{y} = 0; n_{x} =0, n_{y} = 1;
 n_{x} = 1, n_{y} = 0$ and $\rho_{1,1} =
0.$ Having done this we next embark upon the task of finding the
eigenvalues in the regime $ 2 < t' < \infty$. To find an explicit form
of $f'(t')$ in this region we note that in the limiting cases $t'=2$
and $t'=\infty$ the semi-empirical formula should yield,
\begin{equation}
\omega(2)= \omega_{c} \ \ \mathrm{and} \ \ \omega(\infty)= \omega_{s},
\end{equation}
i.e. we require,
\begin{equation}
f'(2)=0  \ \ \mathrm{and} \ \ f'(\infty)=1.
\end{equation}
This is where the duality relation ({\bf \ref{dual}}) comes into play.
Plugging it in ({\bf \ref{f}}) we arrive at the
following functional form of $f'(t')$,
\begin{equation}
f'(t')=\left(\frac{1-\frac{2}{t'}}{1-\frac{1}{t'}}\right)^{3}.
\end{equation}
Clearly, such a choice of $f'(t')$ satisfies the limiting conditions.
So, finally for $t'$ lying in the range $2 < t' < \infty$ the duality
implies that the expression for the eigenvalues will be:
\begin{align}
\omega(t')=\left[\left(\frac{1-\frac{2}{t'}}{1-\frac{1}{t'}}\right)^{3}{\cal{E}}_{s}(n_x,n_y)+\left\{1-\left(\frac{1-\frac{2}{t'}}{1-\frac{1}{t'}}\right)^{3}\right\}{\cal{E}}_{c}(\rho_{l,j})\right]\frac{t'\Gamma(\frac{2}{t'})}{2[\Gamma(\frac{1}{t'})]^{2}}. \label{second}
\end{align}
Using the above expressions we have evaluated the eigenvalues for the
first 21 states (including the degenerate ones) for a supercircle and
compared them with the numerically obtained values. The comparisons
are shown in FIG.\ref{fig:VIII}, FIG.\ref{fig:IX} and FIG.\ref{fig:X}.  We
find that the agreement is excellent over a wide range of the
supercircular exponent $t$
except for a few cases which we discuss in the next section.
\subsection{Neumann boundary condition}
\begin{figure}
\centering
\psfrag{Energy (E)}[c][c][1][0]{{\bf {\LARGE Energy ($\omega$)}}}
\rotatebox{0}{\scalebox{0.6}{\includegraphics{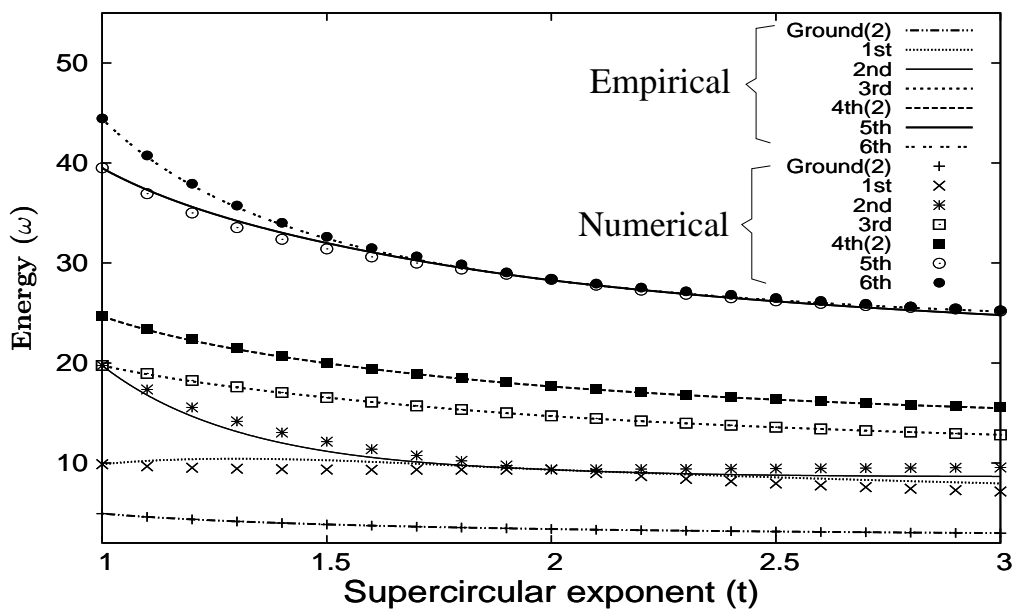}}}
\caption{Comparison of the eigenvalues obtained numerically
  and empirically for a supercircular boundary with Neumann condition
  for the first 9 states.}
\label{fig:XI}
\end{figure}

\begin{figure}
\centering
\psfrag{Energy (E)}[c][c][1][0]{{\bf {\LARGE Energy ($\omega$)}}}
\rotatebox{0}{\scalebox{0.6}{\includegraphics{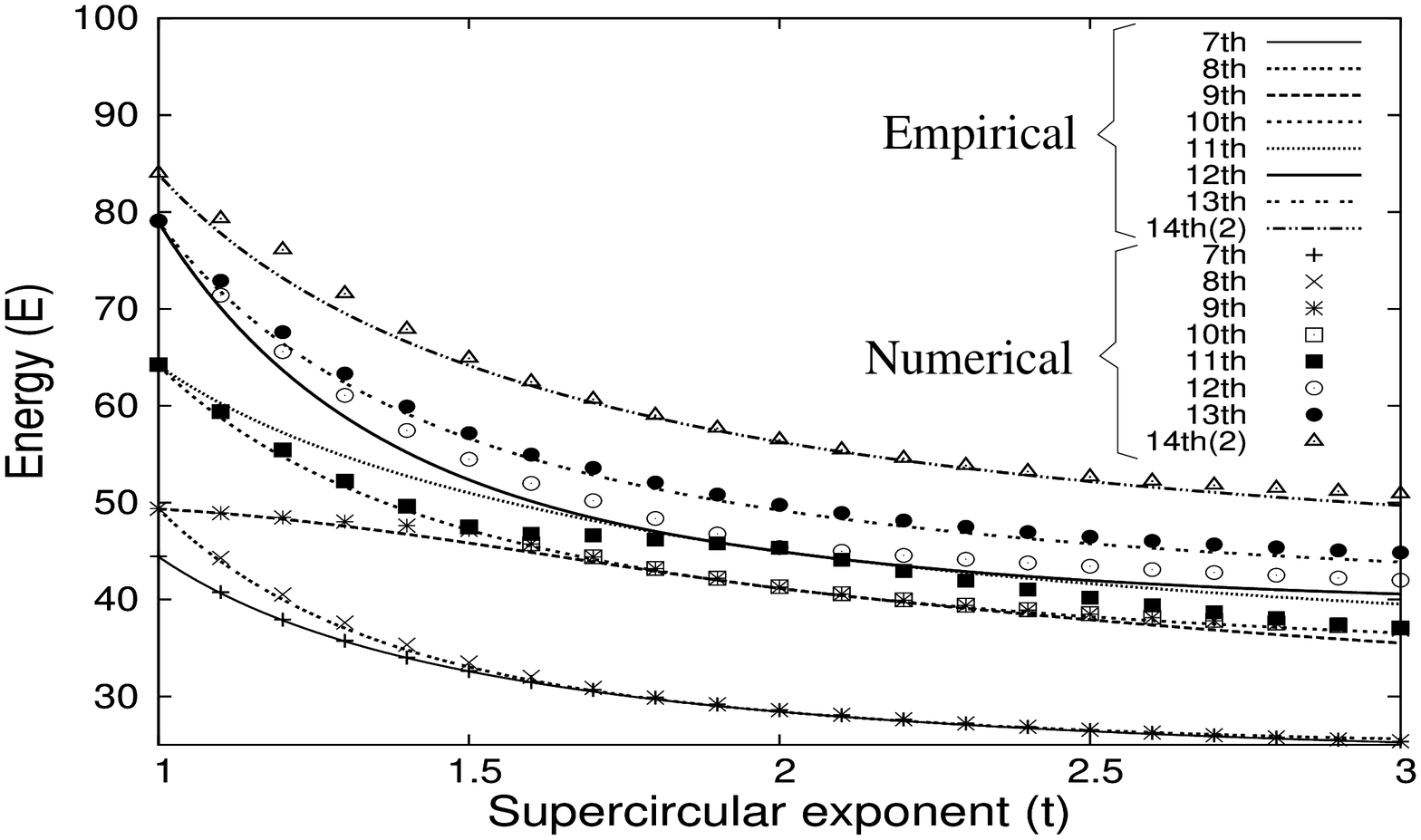}}}
\caption{Comparison of the eigenvalues obtained numerically
  and empirically for a supercircular boundary with Neumann condition
  for the states 10 to 18.}
\label{fig:XII}
\end{figure}

In case of the Neumann boundary condition we must have $\frac{\partial
  \psi}{\partial n}=0$ on $\partial \Sigma$.  The vanishing of the
normal derivative implies that for the square boundary with side $L$
the eigenfunction should now be,
\begin{equation}
  \psi= \frac{2}{L}\cos\left(\frac{n_{x}\pi
    x}{L}\right)\cos\left(\frac{n_{y}
\pi y}{L}\right).
 \end{equation}
This, however, has no effect on the eigenvalues with the exception of
$n_x=0$, $n_y=1$ and $n_x=1$, $n_y=0$ for which one now has non
trivial $\psi$. On the other hand, for the circular boundary of radius
$R$ the eigenfunction remains the same as before, but the Neumann
boundary condition dictates that we must have,
\begin{equation}
\omega_{c} \sim k_{c}^{2}=\frac{{\rho'}_{l,j}^{2}}{R^{2}}
\end{equation}
where $\rho'_{l,j}$ is now the $j^{th}$ zero of the derivative of the
$l^{th}$ order Bessel function, i.e. $J_l'(\rho'_{l,j})=0$. Barring
these slight modifications, the basic recipe for finding out the
eigenvalues remains unaltered from the Dirichlet case.  The
eigenvalues are given by the expressions ({\bf \ref{first}}) and ({\bf
  \ref{second}}) with properly matched ${\cal E}_s$ and ${\cal
  E}_c$. In Table \ref{table:2} we have illustrated the exact matching
between the parameters by considering some low lying modes. Here also
we have excluded the trivial case $n_{x} = n_{y} = 0 $ and
$\rho'_{0,1} = \rho_{1,1} = 0$ as it was done for the Dirichlet
case. We have plotted the eigenvalues for the first 18 states
(including the degenerate ones) and compared them with the numerical
results in FIG.\ref{fig:XI} and FIG.\ref{fig:XII}.  Once again the
matching is found to be reasonably good 
save and except a few levels whose discussion we postpone to the next section.
Comparison of the empirical and the numerical eigenvalues
for the first seven states for a supercircle with exponent $t=1.5$ for
both type of boundary conditions is given in Table \ref{table:3}.
\section{CONCLUSION} 
In this paper we have generalised the formalism developed in
\cite{chak} to encompass the case of Neumann type boundary condition.
We have also explicitly verified the results for two boundary
geometries - supercircle and ellipse. This is quite significant since
the Helmholtz equation is not exactly solvable for such geometries.
Here we briefly highlight the salient features which make our
formalism different from the existing ones.  The scheme we develop is
completely different from those prevalent to attack such kind of
boundary value problems. However, it does have a parallel in the
Rayleigh-Schr\"{o}dinger type potential perturbation in quantum
mechanics.  In fact, one of the primary motivations of the order by
order construction of the Fourier series in \cite{chak} was to make
contact with the time independent perturbation theory in quantum
mechanics.  In 2-level systems in quantum mechanics one witnesses the
phenomena of level crossing when the Hamiltonian is perturbed.  This
is most easily achieved by placing the system in a constant electric
or magnetic field \cite{Gottfried}. The analogy with quantum mechanics
is further consolidated with the observation that boundary
perturbation can also induce crossing of the levels. Although in
quantum mechanics one usually solves a Dirichlet boundary value
problem, here we show that the phenomena of level crossing can appear
even in Neumann type problems. Another unique feature is that unlike
any other perturbation scheme ours does not require a separate
formalism to treat degeneracies.  Degenerate states are handled
equally elegantly as the non-degenerate ones. As opposed to earlier
schemes where the perturbation parameter is an artificial one here a
real one, extracted from the parameters defining the irregular
boundary, serves as a perturbation parameter. It is found that for any
perturbation the higher states are more affected. We have plotted
beyond the 20th excited level and have shown that the analytical
perturbation procedure developed here does reasonably well even for
such higher states. Finally, our scheme is a one shot process.  If
there is an equation for the family of closed curves (like that in an
ellipse or a supercircle) one can have the full spectrum just from a
knowledge of the parameters of the equation. We exploit the parameters
of curve equation as the perturbation parameter which is a measure of
the deviation from a circle and obtain the results in one go. One has
to just substitute the parameter values to get the new spectrum
whereas in all other existing schemes one has to find each time the
Fourier coefficients and then substitute even if the curve is of the
same family.

Besides the phenomenon of level crossing,
another interesting case is when the
levels just manage to evade a crossing and veers off after touching
each other tangentially.  For the 2-level quantum mechanical system 
mentioned here one can
in fact set up a criteria to predict whether or not a pair of levels
will cross. It will be interesting to explore the possibility of
formulating an analogous criteria for boundary perturbation. A closely
related problem is to investigate why the perturbation theory falters
whenever the levels display repulsion for elliptical boundary.
We have seen that for the elliptical boundary,
the agreement between the analytical
and numerical results is quite outstanding for small values of
$\lambda$. The only blemish is the inability of the scheme to predict
the level repulsions.  A proper understanding of this level repulsion
can be a potential topic for further study.

We have also put forward a single parameter semi-empirical formula
giving the eigenspectrum of the 2 dimensional homogeneous Helmholtz
equation with Dirichlet or Neumann condition on a supercircular
boundary. It is found that the prescription gives results to a high
degree of accuracy for a wide range of the parameter $t$ for both type
of boundary conditions barring a few cases. In fact we have checked
explicitly for $1< t < 5$ and the agreement is rather impressive
throughout the range.  This is quite remarkable considering the fact
that a semi-empirical formula involving just a single parameter, the
supercircular exponent $t$, yields such accurate results. This assumes
more significance because the problem is not exactly solvable
analytically. Even the perturbative scheme that we have developed
works only for small deformations about the circle.  On the other
hand, the semi empirical formula, supplemented with the duality
relation gives us an analytic handle to extract the eigenvalues even
when $ t \gg 2$.  A possible generalisation would be to attempt a
similar formula when $t \leq 1$, i.e. we have a concave curve.  We
would also like to point out a very peculiar feature that has plagued
the semi-empirical results: the abnormal behaviour
of the Bessel function $J_{2}$. The semi-empirical formula for the
Dirichlet boundary condition does not provide the anticipated results
whenever there is a contribution to $\omega_{c}$ from the zeros of
$J_2$ which is quite mysterious indeed. The first such case arises
from $\rho_{2,1}$ which contributes to the eigenvalues of the second
and the third modes. It is clear from Table \ref{table:3} that for the
Dirichlet case the agreement between the empirical and the numerical
values is outstanding and the maximum error is $0.6\%$ except for the
second and the third states where the error is within $4\%$. The next
victims are the ninth and the tenth modes which receive contribution
from the second zero $\rho_{2,2}$ of $J_2$ and in both cases the
degree of mismatch is little pronounced but the relative errors are
within $5\%$. In the case of the Neumann condition also there is
disparity between the numerical and the semiempirical predictions for
some of the eigenvalues. The first instance occurs in the case of the
first and the second modes both of which involve a contribution from
$\rho'_{2,1}$. Table \ref{table:3} also shows that the errors for the
aforesaid modes are around $9\%$ for $t=1.5$. For the other modes the
matching is very good and the errors are within $2\%$. The next levels
afflicted with this anomaly are the eleventh and the twelfth ones
which involve $\rho'_{2,2}$.  This mysterious behaviour of $J_{2}$ is
something which cannot be accounted for at this point and calls for
deeper study.  Strangely enough, all these  $J_{2}$ states  also involve
level repulsion.  In this context, it may be mentioned that
Chakraborty \textit{et al} \cite{chak} had earlier reported the failure of
perturbative treatment in the cases where the levels repel each other
for the case of Dirichlet condition on an elliptical boundary and
there also the levels were associated with $J_2$.  Here too the
analytical scheme for the Neumann boundary on the supercircle 
reveals that the $J_{2}$ states display level repulsion.
From the results at hand it seems as if level repulsion and $J_{2}$
abnormality go hand in hand. But we do not know whether this is sheer
coincidence or there is an underlying connection between the two. In
any case we think it would be worthwhile to look into this aspect.

\section{Acknowledgments}
SP would like to acknowledge Council of Scientific and Industrial
Research (CSIR), India for providing the financial support.


\clearpage

\begin{center}
\begin{table}
  \begin{tabular}{|c|c|c|c|c|c|c|c|}
  \hline
  $n_x$   & $n_y$  & ${\cal E}_s/\pi^2 $ & State 
&{\phantom x}$l${\phantom x} &{\phantom x}$j${\phantom x} 
&$\rho_{l,j}$ & ${\cal E}_c/\pi$      \\
          &        & $= (n_x^2 + n_y^2 )$     &          
&    &   &&= $\rho^2_{l,j}$\\ \hline 
  1 & 1 & 2 & Ground    & 0 & 1 &2.40483& 5.7832 \\  \hline
  1 & 2 & 5 & $1^{st}$   & 1 & 2 &3.83171&14.6820 \\ 
  2 & 1 & 5 &           & 1 & 2 &3.83171& 14.6820 \\ \hline
  2 & 2 & 8 & $2^{nd}$   & 2 & 1 &5.13562& 26.3746 \\  \hline
  1 & 3 & 10 & $3^{rd}$  & 2 & 1 &5.13562& 26.3746 \\  \hline
  3 & 1 & 10 & $4^{th}$  & 0 & 2 &5.52008& 30.4713 \\  \hline
  2 & 3 & 13 & $5^{th}$  & 3 & 1 &6.38016&40.7064 \\ 
  3 & 2 & 13 &   & 3 & 1 &6.38016& 40.7064 \\ \hline
  1 & 4 & 17 & $6^{th}$  & 1 & 3 &7.01559& 49.2185 \\ 
  4 & 1 & 17 &   & 1 & 3 &7.01559& 49.2185 \\ \hline
  \end{tabular}

\caption{Matching of the parameters of ${\cal E}_s$ and ${\cal E}_c$
  for 
finding $\omega(t)$ for the Dirichlet boundary condition}
\label{table:1}
\end{table}
\end{center}

\begin{center}
\begin{table}
  \begin{tabular}{|c|c|c|c|c|c|c|c|}
  \hline
   $n_x$   & $n_y$  & ${\cal E}_s/\pi^2 $     & State    
& {\phantom x}$l${\phantom x} &{\phantom x}$j${\phantom x} 
&$\rho'_{l,j}$ & ${\cal E}_c/\pi$      \\
           &        & $= (n_x^2 + n_y^2 )$     &          &     &
   &            
&= ${\rho'}^2_{l,j}$               \\ \hline 

  0 & 1 & 1 & Ground & 1 & 1 & 1.8412&3.3900 \\ 1 & 0 & 1 & & 1 & 1 &
  1.8412&3.3900 \\ \hline 1 & 1 & 2 & $1^{st}$ & 2 & 1 & 3.0542&9.3281
  \\ \hline 0 & 2 & 4 & $2^{nd}$ & 2 & 1 & 3.0542&9.3281 \\ \hline 2 &
  0 & 4 & $3^{rd}$ & 0 & 2 & 3.8317&14.6819 \\ \hline 1 & 2 & 5 &
  $4^{th}$ & 3 & 1 & 4.2012&17.6501 \\ 2 & 1 & 5 & & 3 & 1 &
  4.2012&17.6501 \\ \hline 2 & 2 & 8 & $5^{th}$ & 4 & 1 &
  5.3175&28.2758 \\ \hline 0 & 3 & 9 & $6^{th}$ & 4 & 1 &
  5.3175&28.2758 \\ \hline
  \end{tabular}
\caption{Matching of the parameters of ${\cal E}_s$ and ${\cal E}_c$
  for finding $\omega(t)$ for the Neumann boundary condition}
\label{table:2}
\end{table}
\end{center}

\begin{center}
\begin{table}
  \begin{tabular}{|c|c|c|c|c|c|c|}
  \hline
     State    & \multicolumn{3}{c|}{Dirichlet condition}  
&  \multicolumn{3}{c|}{Neumann condition}           \\
\cline{2-7}
    & Degeneracy & Empirical $\omega(1.5)$ & Numerical $\omega(1.5)$ 
& Degeneracy &Empirical $\omega(1.5)$ & Numerical $\omega(1.5)$     \\ 
\hline 
Ground&1&6.7077&6.7178&2&3.8542&3.8477\\ \hline
First &2&16.9942&16.9954&1&10.2669&9.3361\\ \hline
Second&1&30.0858&29.0001&1&11.1682&12.1291\\ \hline
Third &1&30.9870&32.0603&1&16.5436&16.5417\\ \hline
Fourth&1&35.1002&35.0087&2&19.9743&20.0064\\ \hline
Fifth &2&46.7285&46.4754&1&31.9947&31.4117\\ \hline
Sixth &2&57.0773&57.1111&1&32.4453&32.6240\\ \hline
  \end{tabular}
\caption{Comparison between empirically and numerically calculated 
eigenvalues for $t = 1.5$}
\label{table:3}
\end{table}
\end{center}







\begin{thebibliography}{}
\bibitem{APL} I. Sobchenko, J. Pesicka, D.Baither, R. Reichelt, E. Nembach,
Applied Physics Letters \textbf{89}, 133107 (2006).
\bibitem{Bera} N. Bera, J.K. Bhattacharjee, S. Mitra, S.P. Khastgir,
Eur. Phys. J. D \textbf{46}, 41 (2008).
\bibitem{chak} S. Chakraborty, J.K. Bhattacharjee, S.P. Khastgir,
J. Phys. A: Math. Gen. \textbf{42}, 195301 (2009).
\bibitem{Lis} K. Lis, S. Bednarek, B. Szafran, J. Adamowski,
Physica E \textbf{17}, 494 (2003).
\bibitem{Drouvelis} P.S. Drouvelis, P. Schmelcher, F.K. Diakonos,
Physical Review B \textbf{69}, 155312 (2004).
\bibitem{Prb} I. Magn\'{u}sd\'{o}ttir, V. Gudmundsson,
Physical Review B \textbf{60}, 16591 (1999).
\bibitem{Kac} M. Kac,
American Mathematical Monthly \textbf{73}, 1-23 (1966).
\bibitem{Gordon} C. Gordon, D. Webb, S. Wolpert,
Inventiones Mathematicae \textbf{110}, 1 (1992).
\bibitem{Doron} E. Doron, U. Smilansky,
Nonlinearity \textbf{5}, 1055 (1992).
\bibitem{Dietz} B. Dietz, U. Smilansky,
CHAOS \textbf{3}, 581 (1993).
\bibitem{Berry} M.V. Berry,
J. Phys. A: Math. Gen. \textbf{27}, L391 (1994).
\bibitem{Eckmann} J.P. Eckmann, C.A. Pillet, 
Commun. Math. Phys. \textbf{170}, 283 (1995).
\bibitem{Tasaki} S. Tasaki, T. Harayama, A. Shudo, 
Phys. Rev. E \textbf{56}, R13 (1997).
\bibitem {Revisited} Y. Okada, A. Shudo, S. Tasaki, T. Harayama,
J. Phys. A: Math. Gen. \textbf{38}, L163 (2005).


\bibitem{Krishnamurthy} H.R. Krishnamurthy, H.S. Mani, H.C. Verma,
J. Phys. A: Math. Gen. \textbf{15}, 2131 (1982).
\bibitem{M1} J. Mazumdar,
Shock and Vibration Digest \textbf{7}, 75  (1975).
\bibitem{M2} J. Mazumdar,
Shock and Vibration Digest \textbf{11}, 25 (1979).
\bibitem{M3}  J. Mazumdar,
Shock and Vibration Digest \textbf{14}, 11 (1982).
\bibitem{Kuttler} J.R. Kuttler, V.G. Sigillito,
SIAM Review \textbf{26}, 163 (1984).
\bibitem{Amore} P. Amore,
J. Phys. A: Math. Theor. \textbf{41}, 265206 (2008).
\bibitem{Amore1} P. Amore,
J. Math. Phys. \textbf{51}, 052105 (2010).
\bibitem{Shaw} R.C.T. George, P.R. Shaw,
J. Acoust. Soc. Am. \textbf{56}, 796 (1974).

\bibitem{Wilson} H.B. Wilson, R.W. Scharstein,
Journal of Engineering Mathematics
\textbf{57}, 1 41 (2007).
\bibitem{Hettich} R. Hettich, E. Haaren, M. Ries, G. Still,
Journal of Applied Mathematics and Mechanics
  \textbf{67}, 12 589 (1987).
\bibitem{Troesch} B.A. Troesch, H.R. Troesch,
Mathematics of Computation \textbf{27}, 24 (1973).
\bibitem {Kaufman} D.L. Kaufman, I. Kosztin,  K. Schulten,
Am. J. Phys. \textbf{67}, 133 (1998).
\bibitem {Vergini} E. Vergini, M. Saraceno,
Physical Review B \textbf{52}, 2204 (1995).
\bibitem {Cohen} D. Cohen, N. Lepore, E.J. Heller,
J. Phys. A: Math. Theor. \textbf{37}, 2139 (2004).
\bibitem {Kosztin} I. Kosztin, K. Schulten,
Int. J. Mod. Phys. C \textbf{8}, 233 (1997).
\bibitem {Robnik} M. Robnik,
 J. Phys. A: Math. Theor. \textbf{17}, 1049 (1984).
\bibitem {Erwin} E. Lijnen, L.F. Chibotaru, A. Ceulemans,
Physical Review B \textbf{77}, 016702 (2008).
\bibitem{Rayleigh} J.W.S.B. Rayleigh, \textit{Theory of Sound},
2nd. ed., (Dover, New York, 1945).
\bibitem{Fetter} A.L. Fetter, J.D. Walecka, \textit{Theoretical Mechanics
  of Particles and Continua}, (McGraw Hill Book Company, 1980).
\bibitem{Morse} P.M. Morse, H. Feshbach, \textit{Methods of Theoretical
  Physics}, (Vol.2, McGraw Hill Book Company, 1983).
\bibitem{Parker} R.G. Parker, C.D. Mote Jr., 
Journal of Sound and
  Vibration  \textbf{211}, 3 389 (1998).
\bibitem{Nayfeh} A.H. Nayfeh, \textit{Introduction to Perturbation
  Techniques}, (J. Wiley, New York, 1981).
\bibitem{Read} W.W. Read,
 Mathematical and Computer Modelling
\textbf{24}, 2 23 (1996).
\bibitem{Shiva} Y. Wu, P.N. Shivakumar,
Computers and
  Mathematics with Applications \textbf{55}, 6 1129 (2008).
\bibitem{Molinari} L. Molinari,
J. Phys. A. Math. Gen. \textbf{30}, 6517 (1997).
\bibitem{Dub} R. Dubertrand, E. Bogomolny, N. Djellali, M. Lebental,
  C. Schmit, Physical Review A \textbf{77}, 013804 (2008).
\bibitem{Gardner} M. Gardner, Piet Hein's Superellipse, Ch. 18 in
\textit{Mathematical Carnival: A new Round-Up of Tantalizers and
Puzzles from Scientific American.}, (New York: Vintage, pp.
240-254, 1977).
\bibitem{Gridgeman} N.T. Gridgeman, Lam\'{e} Ovals, Math. Gaz.
\textbf{54}, 31 (1970).
\bibitem{Gottfried} K. Gottfried, T. Yan,  \textit{Quantum Mechanics:
  Fundamentals}, 2nd ed. (Springer, 2003).
\end{thebibliography}
\end {document}